\documentclass[11pt,a4paper]{article}
\pdfoutput=1
\usepackage{jheppub}
\usepackage[utf8]{inputenc}

\usepackage{bbm}
\usepackage{ifthen}

\renewcommand\phi\varphi

\DeclareMathOperator\Trace{tr}
\DeclareMathOperator\sign{sgn}
\DeclareMathOperator{\re}{Re}
\DeclareMathOperator{\im}{Im}

\newcommand{\I}[1]{{\mathcal H}^s_{\nu,#1}(\ha,\hm)} 

\newcommand\ev[1]{\left\langle{#1}\right\rangle}
\newcommand{\ha}{\hat \alpha}
\newcommand{\hm}{\hat m}
\newcommand{\hz}{\hat z}
\newcommand{\hp}{p_s}
\newcommand{\hr}{r_s}
\newcommand{\hw}{w_s}
\newcommand{\ph}{\bigl\langle e^{2i\theta} \bigr\rangle}
\newcommand{\phs}{\bigl\langle e^{2i\theta}_s \bigr\rangle}

\newcommand{\Nf}{{N_f}}

\renewcommand{\epsilon}{\varepsilon}
\newcommand{\partp}[3]
{
\ifthenelse{\equal{#1}{1}}{\partial_m}{\partial_m^{#1}}
p_{#2}^{\nu,#3}(m;\alpha)
}

\title{The QCD sign problem and dynamical simulations of random matrices}

\author{Jacques Bloch and Tilo Wettig}
\affiliation{Institute for Theoretical Physics, University of Regensburg, 93040
  Regensburg, Germany}
\emailAdd{jacques.bloch@physik.uni-regensburg.de}
\emailAdd{tilo.wettig@physik.uni-regensburg.de}

\subheader{\hfill\normalsize March 1, 2011}

\abstract{ At nonzero quark chemical potential dynamical lattice
  simulations of QCD are hindered by the sign problem caused by the
  complex fermion determinant.  The severity of the sign problem can
  be assessed by the average phase of the fermion determinant.  In an
  earlier paper we derived a formula for the microscopic limit of the
  average phase for general topology using chiral random matrix
  theory. In the current paper we present an alternative derivation of
  the same quantity, leading to a simpler expression which is also
  calculable for finite-sized matrices, away from the microscopic
  limit.  We explicitly prove the equivalence of the old and new
  results in the microscopic limit.  The results for finite-sized
  matrices illustrate the convergence towards the microscopic limit.
  We compare the analytical results with dynamical random matrix
  simulations, where various reweighting methods are used to
  circumvent the sign problem.  We discuss the pros and cons of these
  reweighting methods. }


\begin{document}

\maketitle
\flushbottom

\section{Introduction}

In dynamical lattice simulations of QCD at nonzero quark chemical
potential $\mu$ the generation of a Markov chain through importance
sampling is hindered by the sign problem caused by the complex fermion
determinant, see \cite{deForcrand:2010ys} for a review. The severity
of the sign problem grows as $\mu$ increases and the determinant
fluctuates more strongly.  In the $\epsilon$-regime of QCD, i.e., to
leading order in the $\epsilon$-expansion of chiral perturbation
theory \cite{Gasser:1987ah}, the spectral properties of the Dirac
operator are universal and can be computed in the microscopic limit of
chiral random matrix theory (chRMT)
\cite{Shuryak:1992pi,Verbaarschot:2000dy,Basile:2007ki}. This
equivalence also holds at $\mu\ne0$ so that chRMT can be used as a
tool to investigate the sign problem.
  
The fluctuating behavior of the fermion determinant can be
characterized by is its average phase. Using chRMT, Splittorff and
Verbaarschot have computed the average phase at $\mu\ne0$ for trivial
topology in the quenched and unquenched case \cite{Splittorff:2007ck}.
Their results were later extended to nonzero temperature \cite
{Han:2008xj} and to general topology \cite{Bloch:2008cf}. The complex
analysis employed in ref.~\cite{Bloch:2008cf} is quite involved, and
in the present work we give an alternative derivation of the formula
for general topology, based on ideas presented in
ref.~\cite{Osborn:2008jp}.  Although the final integral expressions
for the microscopic limit of the average phase look quite different in
both cases, we show that they are indeed equivalent. En-passant we
also derive some interesting new integral identities.  In addition,
the new derivation also provides an analytical expression for the
average phase of finite-sized matrices, away from the microscopic
limit. This allows us to verify the analytical formulas numerically
using dynamical chRMT simulations. Such simulations are very costly
and can only be performed with high statistical accuracy for
small-sized matrices. In our dynamical chRMT simulations the complex
weights are implemented using various reweighting methods. We provide
a discussion of the pros and cons of these methods.

The structure of this paper is as follows. In section~\ref{Sec:RMT} we
introduce the chiral random matrix model with a chemical potential.
In section~\ref{Sec:Phase} we show how the average phase of the fermion
determinant can be computed in this model using complex Cauchy
transforms.  In section~\ref{Sec:Cauchy} the complex Cauchy transform is
solved for finite-sized matrices, and in section~\ref{sec:micro} the
microscopic limit is taken. In section~\ref{sec:equiv} we prove the
equivalence of the integral representations of the microscopic limits
derived here and in ref.~\cite{Bloch:2008cf}.  In
section~\ref{sec:results} we verify the analytical predictions for the
unquenched case by random matrix simulations away from the microscopic
limit, using different reweighting methods. Finally we draw
conclusions in section~\ref{Sec:concl}.  Intermediate steps of the
calculations are worked out in several appendices.

\section{Random matrix model}
\label{Sec:RMT}

Throughout this paper we use the same conventions as in
ref.~\cite{Bloch:2008cf}.  To make the presentation self-contained, we
reproduce some of the equations derived in that paper.  Details
omitted here can be found in \cite{Bloch:2008cf}.

We work with the non-Hermitian chiral random matrix model for the
Dirac operator $D$ in the presence of a quark chemical potential
introduced by Osborn \cite{Osborn:2004rf},
\begin{equation} D(\mu) =
  \begin{pmatrix}
    0 & i\phi_1 + \mu \phi_2 \\
    i \phi_1^\dagger + \mu \phi_2^\dagger & 0
  \end{pmatrix},
  \label{Dran}
\end{equation}
where the matrices $\phi_1$ and $\phi_2$ are complex random matrices
of dimension $(N+\nu) \times N$.  They are distributed according to a
Gaussian weight function given by
\begin{equation} 
  w(X) = (N/\pi)^{N(N+\nu)} \exp(-N \Trace X^\dagger X)\, .
\label{rmtdis}
\end{equation} 
For a detailed analysis of this model, see also
ref.~\cite{Akemann:2004dr}.  
Since the matrix in eq.~\eqref{Dran} has $|\nu|$ exact zero modes we
can identify $\nu$ with the topological charge.  From now on we
assume $\nu\ge0$; the results for
$\nu<0$ follow by the replacement $\nu\to|\nu|$ in the final results.
The nonzero eigenvalues of $D(\mu)$ come in $N$ pairs $(z_k,-z_k)$,
and for $\mu=0$ the $z_k$ are purely imaginary.  Note that
$\mu$ in \eqref{Dran} is a dimensionless random matrix
quantity and should not be confused with the physical chemical
potential, see the beginning of section~\ref{sec:micro}.

For given $\nu$, the partition function of the random matrix model
with $N_f$ dynamical quarks with masses $m_f$ is
\begin{equation}
  Z_\nu^{N_f}(\mu;\{m_f\}) = \int d\phi_1 d\phi_2 \, w(\phi_1) w(\phi_2)
  \prod_{f=1}^{N_f}\det(D(\mu)+m_f)
  \label{partfun} 
\end{equation}
with integration measure
\begin{equation}
  dX=\prod_{k=1}^{N+\nu}\prod_{\ell=1}^N d\re X_{k\ell} \: d\im X_{k\ell}\,.
\end{equation}
The quenched case corresponds to $N_f=0$.

In ref.~\cite{Osborn:2004rf} it was shown that the partition function
can be rewritten, up to a normalization constant, as an integral over
the eigenvalues $z_k$ of $D$,
\begin{align}
  Z_\nu^{N_f}(\alpha;\{m_f\}) = \int_{\mathbb{C}^N} \Bigg[ \prod_{k=1}^{N}  d^2 z_k \: 
  w^\nu(z_k, z_k^*;\alpha) \prod_{f=1}^{N_f}
  (m_f^2-z_k^2) \Bigg]  |\Delta_N(\{z^2\})|^2\,,
  \label{eq:ZNf}
\end{align}
where we introduced $\alpha = \mu^2$, the integrals over the $z_k$ are
over the entire complex plane,
\begin{align}
  \Delta_N(\{z^2\}) = \prod_{k>\ell} (z_k^2-z_\ell^2) 
\end{align}
is a Vandermonde determinant, the weight function is given by
\begin{equation}
  w^\nu(z,z^*;\alpha)  = |z|^{2\nu+2}
  \exp\left(-\frac{N(1-\alpha)}{4\alpha}(z^2+{z^*}^2)\right) 
  K_\nu\left(\frac{N(1+\alpha)}{2\alpha} |z|^2\right),
\label{wnu}
\end{equation} 
and $K_\nu$ is a modified Bessel function.  The orthogonal polynomials
corresponding to the weight function \eqref{wnu} are
\cite{Osborn:2004rf}
\begin{align}
  p_k^\nu(z;\alpha) = \left(\frac{1-\alpha }{N}\right)^k k! \:
  L_k^\nu\left(-\frac{N z^2}{1-\alpha }\right) \,, 
\label{pknu}
\end{align}
where $L_k^\nu(z)$ is the generalized Laguerre polynomial of order
$\nu$ and degree $k$. The corresponding orthogonality relation is
\begin{align}
 \int_\mathbb{C} d^2z \: w^\nu(z,z^*;\alpha) p_k^\nu(z;\alpha)
 p_\ell^\nu(z;\alpha)^* = r_k^\nu(\alpha) \delta_{k\ell}  
\label{portho}
\end{align}
with norm
\begin{align}
  r_k^\nu(\alpha) = \frac{\pi\alpha(1+\alpha)^{2k+\nu}k!(k+\nu)!} 
  {N^{2k+\nu+2}} \,. 
\label{rknu}
\end{align}
The recurrence relation for the generalized Laguerre polynomials,
\begin{equation}
  (k+1) L^\nu_{k+1}(x) = (k+1+\nu) L^\nu_k(x) - x L^{\nu+1}_{k}(x) \,,
  \label{Lagrecur}
\end{equation}
translates into a recurrence relation for the orthogonal polynomials
$p_k^\nu$, 
\begin{align}
  p_{k+1}^\nu(z;\alpha) &= (k+1+\nu) \left(\frac{1-\alpha }{N}\right)
  p_k^\nu(z;\alpha) + z^2 p_k^{\nu+1}(z;\alpha) \,.
  \label{pkrecur}
\end{align}
We will also use the Cauchy transform of the orthogonal polynomials
defined by
\begin{align}
  h_k^\nu(m;\alpha) = \int_\mathbb{C} \frac{d^2z}{z^2-m^2}
  w^\nu(z,z^*;\alpha)  p_k^\nu(z;\alpha)^* \,. 
\label{hknu}
\end{align}

The ensemble average of an observable $\mathcal O$ is given by
\begin{equation}
  \label{eq:qav}
  \ev{\mathcal O}_{\nu,N_f}=\frac1{Z_\nu^{N_f}}
  \int_{\mathbb{C}^N} \Bigg[ \prod_{k=1}^{N}  d^2 z_k \:      
  w^\nu(z_k, z_k^*;\alpha)  \prod_{f=1}^{N_f}
  (m_f^2-z_k^2) \Bigg] |\Delta_N(\{z^2\})|^2 \: \mathcal O(z_1,\ldots,z_N) \,.
\end{equation}
We will frequently omit one or both of the subscripts on $\ev{\mathcal
  O}$.

\section{Average phase of the fermion determinant} 
\label{Sec:Phase}

Adding a quark mass $m$ to the Dirac operator we define $D(m;\mu) =
D(\mu) + m \mathbbm{1}$, where $m$ is assumed to be real.  Writing
$\det D(m;\mu) = r e^{i\theta}$, the phase of the determinant follows
from \cite{Splittorff:2007ck}
\begin{align}
  \label{eq:ph}
  e^{2i\theta} = \frac{\det(D(\mu)+m)}{\det(D^\dagger(\mu)+m)} 
  =  \prod_{k=1}^{N} \frac{m^2-z_k^2}{m^2-{z_k^*}^2} \,.
\end{align}
Here, $m$ is viewed as a valence quark mass.  We are interested in the
ensemble average of $e^{2i\theta}$ with two light dynamical quarks
that have the same mass as the valence quark.  This quantity is a
measure of the fluctuations of the two-flavor determinant in the QCD
weight function.  For brevity we call $e^{2i\theta}$ the phase of the
determinant, although it really is the phase of the two-flavor
determinant.  Note that the average phase is real since each matrix
appears in the ensemble with the same probability \eqref{rmtdis} as its
Hermitian conjugate.

In the presence of $N_f$ dynamical quarks the average phase for a
valence quark of mass $m$ is given by
\begin{align}
  \ph_{N_f}
  &=\left\langle
    \frac{\det(D(\mu)+m)}{\det(D^\dagger(\mu)+m)}\right\rangle_{\!\!N_f}
  \notag\\
  &= \frac{1}{Z^\Nf_\nu(\alpha;\{m_f\})}\int_{\mathbb{C}^N} \!
  \Bigg[ \prod_{k=1}^{N} d^2 z_k \,  
  w^\nu(z_k, z_k^*;\alpha) \, 
  \frac{m^2-z_k^2}{m^2-{z_k^*}^2} \prod_{f=1}^\Nf (m_f^2-z_k^2) \Bigg] |\Delta_N(\{z^2\})|^2 \notag\\
  &= \frac{Z^{N_f+1|1^*}_\nu\!\!(\alpha,m;\{m_f\})}
  {Z^{N_f}_\nu(\alpha;\{m_f\})} \,,
  \label{unqphfac}
\end{align}
where $Z^\Nf_\nu$ is given by eq.~\eqref{eq:ZNf} and
$Z_\nu^{N_f+1|1^*}$ is the partition function of a random matrix model
with $N_f+1$ fermionic quarks and one conjugate bosonic quark, see
ref.~\cite{Splittorff:2007ck} for a detailed discussion.  Both
partition functions can be interpreted, up to a common additional
normalization factor $Z_\nu^0$, as averages of ratios of characteristic
polynomials in the quenched ensemble. Such averages can be computed in
terms of the orthogonal polynomials \eqref{pknu} and their Cauchy
transforms \eqref{hknu} using the formalism developed in
refs.~\cite{Bergere:2004cp,Akemann:2004zu}. The details of its
application to eq.~\eqref{unqphfac} can be found in
ref.~\cite[section~3.1]{Bloch:2008cf}, and one obtains
\begin{align}
  \ph_{N_f}
  &= 
  \frac{
 \begin{vmatrix}
    {\mathcal H}_{\nu,0}(\alpha,m) & {\mathcal H}_{\nu,1}(\alpha,m) & \cdots & {\mathcal H}_{\nu,\Nf+1}(\alpha,m)\\[1mm]
    p_{N-1}^{\nu,0}(m;\alpha) & p_{N-1}^{\nu,1}(m;\alpha) & \cdots & p_{N-1}^{\nu,\Nf+1}(m;\alpha)\\[1mm]
    p_{N-1}^{\nu,0}(m_1;\alpha) & p_{N-1}^{\nu,1}(m_1;\alpha) & \cdots & p_{N-1}^{\nu,\Nf+1}(m_1;\alpha)\\
    \vdots & \vdots & \vdots & \vdots \\
    p_{N-1}^{\nu,0}(m_\Nf;\alpha) & p_{N-1}^{\nu,1}(m_\Nf;\alpha) & \cdots & p_{N-1}^{\nu,\Nf+1}(m_\Nf;\alpha)
  \end{vmatrix}
  }
  {\left[\prod_{f=1}^\Nf (m_f^2-m^2)\right]
    \det\big[p_{N}^{\nu,g-1}(m_{f};\alpha)\big]_{f,g=1,\ldots,N_f}}\,,
  \label{unqphfac2} 
\end{align}
where we introduced the notation 
\begin{align}
p_{\ell}^{\nu,k}(z;\alpha) = z^{2k} p_{\ell}^{\nu+k}(z;\alpha) 
\label{plnuk}
\end{align}
and defined the complex integral
\begin{align}
{\mathcal H}_{\nu,k}(\alpha,m) =
-  \frac{1}{r_{N-1}^\nu(\alpha)} \int_\mathbb{C} \frac{d^2z}{z^2-m^2}
 w^\nu(z,z^*;\alpha)
 p_{N-1}^{\nu,k}(z^*) 
\label{HnukN}
\end{align}
over the orthogonal polynomials.
In the quenched case \eqref{unqphfac2} simplifies to
\begin{align}
  \ph_{N_f=0}
  =     \begin{vmatrix}
    {\mathcal H}_{\nu,0}(\alpha,m) & {\mathcal H}_{\nu,1}(\alpha,m) \\[1mm]
    p_{N-1}^{\nu,0}(m;\alpha) & p_{N-1}^{\nu,1}(m;\alpha)
  \end{vmatrix} \,.
  \label{Z1by1}
\end{align}

We now consider eq.~\eqref{unqphfac2} for the special
case in which all dynamical fermions have the same mass $m$ as the
valence quark. We perform a Taylor
expansion of the entries $p_\ell^{\nu,k}(m_f)$ of the determinant around
$m$,
\begin{align}
p_{\ell}^{\nu,k}(m_f;\alpha)
  = p_{\ell}^{\nu,k}(m;\alpha)
  + \sum_{j=1}^{\infty}
  \frac{(m_f-m)^j}{j!}  \:  \partp{j}{\ell}{k} \,,
  \qquad f=1, \ldots, \Nf\,.
  \label{mpTaylor}
\end{align}
A determinant remains unaltered when linear combinations of its rows
are added to any of the rows.  Therefore for each additional fermion
it is sufficient to keep the next higher-order term in the expansion
\eqref{mpTaylor}. The lower-order terms do not contribute to the
determinant since they are identical to the contribution from one of
the previous fermions, while the higher-order terms can be neglected
since their contribution vanishes for $m_f \to m$. After taking each
fermion mass in turn to $m$, we obtain
\begin{align}
  \ph_{N_f}
  &= 
  \frac{
 \begin{vmatrix}
    {\mathcal H}_{\nu,0}(\alpha,m) & {\mathcal H}_{\nu,1}(\alpha,m) & \cdots & {\mathcal H}_{\nu,\Nf+1}(\alpha,m)\\[1mm]
    p_{N-1}^{\nu,0}(m;\alpha) & p_{N-1}^{\nu,1}(m;\alpha) & \cdots & p_{N-1}^{\nu,\Nf+1}(m;\alpha)\\[1mm]
    \partp{1}{N-1}{0} & \partp{1}{N-1}{1} & \cdots & \partp{1}{N-1}{\Nf+1}\\[1mm]
    \vdots & \vdots & \vdots & \vdots \\
    \partp{N_f}{N-1}{0} & \partp{N_f}{N-1}{1} & \cdots & \partp{N_f}{N-1}{\Nf+1}  \end{vmatrix}
  }
  {(2m)^\Nf \Nf!\det\big[\partial_m^{f-1}p_{N}^{\nu,g-1}(m;\alpha)
    \big]_{f,g=1,\ldots,N_f}}\,,
  \label{phfacNfflavorsN}
\end{align}
which can also be written in the form
\begin{align}
  \ph_{N_f} = \frac{1}{(2m)^\Nf \Nf!} \: \frac{{\mathcal
      W}^{N-1}_{\Nf}(\alpha,m)}{W_\Nf^N(0, 1, \ldots, \Nf-1)} \,,
\label{phappN}
\end{align}
where
\begin{align}
{\mathcal W}^{N-1}_{\Nf}(\alpha,m) = \sum_{k=0}^{\Nf+1} (-)^k {\mathcal H}_{\nu,k}(\alpha,m) W_{\Nf+1}^{N-1}(0,\ldots, k-1, k+1, \ldots, \Nf+1)
\label{numeratorN}
\end{align}
is a sum of Wronskians of order $\Nf+1$ with indices ranging from $0$
to $\Nf+1$, where in each term a different index $k$ is absent.
The Wronskian
\begin{equation}
  W_{n}^\ell(p_\ell^{\nu,k_1}(m;\alpha), \ldots, p_\ell^{\nu,k_n}(m;\alpha))=
  \det\big[\partial_m^{i-1}p_{\ell}^{\nu,k_j}(m;\alpha)\big]_{i,j=1,\ldots,n}
  \label{G.4N}
\end{equation}
in eqs.~\eqref{phappN} and \eqref{numeratorN} is abbreviated by
$W_{n}^\ell(k_1,\ldots,k_n)$.

\section{Solving the complex Cauchy transform}
\label{Sec:Cauchy}

The complex integral \eqref{HnukN} needed in the computation of the
phase factor is strongly oscillating and cannot easily be evaluated
numerically to high accuracy.  In ref.~\cite{Bloch:2008cf} we solved
this integral in the microscopic limit using quite involved complex
analysis. We introduced integration contours which were then deformed
such that the result was composed of contributions from the branch cut
discontinuity and from the singularity of the modified $K$-Bessel
function. The final result was given in terms of well-behaved
one-dimensional integrals plus a short double sum.  In this section we
present a different derivation, which has the additional advantage of
giving a calculable result for finite $N$, away from the microscopic
limit.

 We first explicitly substitute the weight factor \eqref{wnu} in eq.~\eqref{HnukN},
\begin{align}
{\mathcal H}_{\nu,k}(\alpha,m) =
-  \frac{1}{r_{N-1}^\nu(\alpha)} \int_\mathbb{C} \frac{d^2z}{z^2-m^2} |z|^{2(\nu+1)} e^{-a(z^2+{z^*}^2)} K_\nu(b|z|^2)\, p_{N-1}^{\nu,k}(z^*;\alpha) \,,
\label{HnukNb}
\end{align}
where we defined $a = N(1-\alpha)/4\alpha$ and $b =
N(1+\alpha)/2\alpha$. In ref.~\cite{Osborn:2008jp}, Osborn, Splittorff
and Verbaarschot solved the Cauchy transform \eqref{hknu}, which
corresponds to the special case of
$k=0$ in eq.~\eqref{HnukNb}.\footnote{We thank Jac Verbaarschot for drawing our attention to
  the solution of the Cauchy transform in ref.~\cite{Osborn:2008jp}.}
In the following we extend their result to arbitrary positive integer
$k$, as required in eq.~\eqref{unqphfac2}.  Following the derivation
in section~V of ref.~\cite{Osborn:2008jp} we write
\begin{align}
\frac{e^{-az^2}}{z^2-m^2} = \frac{e^{-am^2}}{z^2-m^2} + \frac{e^{-az^2}-e^{-am^2}}{z^2-m^2} 
\label{split}
\end{align}
and decompose eq.~\eqref{HnukNb} accordingly as ${\mathcal H}_{\nu,k}={\mathcal H}_{\nu,k}^{(1)}+{\mathcal H}_{\nu,k}^{(2)}$ with 
\begin{align}
{\mathcal H}_{\nu,k}^{(1)}(\alpha,m) &=
- \frac{e^{-am^2}}{r_{N-1}^\nu(\alpha)} \int_\mathbb{C} \frac{d^2z}{z^2-m^2} |z|^{2(\nu+1)} e^{-a{z^*}^2} K_\nu(b|z|^2)\, p_{N-1}^{\nu,k}(z^*;\alpha) \,,
\label{HnukN1}\\
{\mathcal H}_{\nu,k}^{(2)}(\alpha,m) &=
-  \frac{1}{r_{N-1}^\nu(\alpha)} \int_\mathbb{C} d^2z |z|^{2(\nu+1)} \frac{e^{-az^2}-e^{-am^2}}{z^2-m^2} e^{-a{z^*}^2} K_\nu(b|z|^2)\, p_{N-1}^{\nu,k}(z^*;\alpha) \,.
\label{HnukN2}
\end{align}

For ${\mathcal H}_{\nu,k}^{(1)}$ we consider the cases $|z|<m$ and $|z|>m$ separately and expand the de\-no\-mi\-na\-tor in a geometric series,
\begin{align}
{\mathcal H}_{\nu,k}^{(1)}(\alpha,m) 
&= -  \frac{e^{-am^2} }{r_{N-1}^\nu(\alpha)}
\bigg[
- \sum_{j=0}^\infty \frac{1}{m^{2(j+1)}}\int_{|z|<m} \!\!\!d^2z \,
 |z|^{2(\nu+1)} K_\nu(b|z|^2) \, z^{2j} e^{-a{z^*}^2} 
 p_{N-1}^{\nu,k}(z^*;\alpha)  \notag\\ 
& \qquad + \sum_{j=0}^\infty m^{2j} \int_{|z|>m} \!\!\!d^2z \,
|z|^{2(\nu+1)} K_\nu(b|z|^2) \, \frac{1}{z^{2(j+1)}} e^{-a{z^*}^2} 
p_{N-1}^{\nu,k}(z^*;\alpha)  
\bigg] 
\,.
\label{Hknu1}
\end{align}
The polynomials $p_{\ell}^{\nu,k}(z;\alpha)$ defined in \eqref{plnuk}
are even in their argument $z$ so that we can implicitly define the expansion 
\begin{align}
f(z^2) = e^{-a{z}^2} p_{N-1}^{\nu,k}(z;\alpha) = \sum_{n=0}^\infty a_n z^{2n} \,.
\label{fz}
\end{align}
In polar coordinates the angular part of the integrands in eq.~\eqref{Hknu1} can therefore be written as a product of power series in $z^2$ and ${z^*}^2$.
With $z=r e^{i\theta}$,
the angular integration of such a product can be computed analytically using
\begin{align}
\int_0^{2\pi} d\theta \: z^{2k} {z^*}^{2\ell} = 2\pi r^{4k} \delta_{k\ell} \,.
\end{align}
After angular integration the second integral in eq.~\eqref{Hknu1} vanishes, while the first one gets contributions
\begin{align}
\int_0^{2\pi} d\theta \: z^{2j} f({z^*}^2) 
= \sum_{n=0}^\infty a_n \int_0^{2\pi} d\theta \: z^{2j} {z^*}^{2n}
= 2\pi a_j r^{4j} \,.
\end{align}
Resumming the $a_j$ using eq.~\eqref{fz},
the integral \eqref{Hknu1} can then be written as
\begin{align}
{\mathcal H}_{\nu,k}^{(1)}(\alpha,m)
&= \frac{2\pi e^{-am^2} }{m^2 r_{N-1}^\nu(\alpha)}
\int_0^m dr \,
 r^{2\nu+3} K_\nu(br^2)\, f\left(\frac{r^{4}}{m^{2}}\right)
\notag\\
&= \frac{\pi m^{\nu} e^{-am^2} }{r_{N-1}^\nu(\alpha)}
\int_0^m du \,  
 u^{\nu+1} K_\nu(bmu) e^{-a{u}^2} p_{N-1}^{\nu,k}(u;\alpha)
\,,
\label{HkN1}
\end{align}
where in the last step we have introduced the variable transformation
$u=r^2/m$ and replaced $f(u^2)$ by its explicit expression \eqref{fz}.

We now turn to ${\mathcal H}_{\nu,k}^{(2)}$ given by
eq.~\eqref{HnukN2}. The second term on the RHS of eq.~\eqref{split} is analytic and can be expanded as
\begin{align}
\frac{e^{-az^2}-e^{-am^2}}{z^2-m^2} = e^{-a z^2} \sum_{n=0}^{\infty} d_n p_n^\nu(z;\alpha) \,,
\label{exp_part2}
\end{align}
where the coefficients $d_n$ were computed in 
ref.~\cite{Osborn:2008jp} and are given by
\begin{align}
d_n = -\frac{N^{n+1}}{n! (1-\alpha)^{n+1}} \int_0^{(1-\alpha)^2/4\alpha} dt \,
e^{-m^2Nt/(1-\alpha)} \frac{t^n}{(t+1)^{n+\nu+1}} \,.
\label{dm}
\end{align}
Note that the explicit factor $e^{-a z^2}$ in eq.~\eqref{exp_part2} is such that  the weight of the orthogonal polynomials are retrieved after substituting eq.~\eqref{exp_part2} in eq.~\eqref{HnukN2}. This yields
\begin{align}
{\mathcal H}_{\nu,k}^{(2)}(\alpha,m) = 
- \frac{1}{r_{N-1}^\nu(\alpha)} 
\sum_{n=0}^{\infty} d_n \int_\mathbb{C} d^2z \, w^\nu(z,z^*;\alpha)  
p_n^\nu(z;\alpha)  p_{N-1}^{\nu,k}(z^*;\alpha) \,.
  \label{Hk}
\end{align}
In ref.~\cite{Osborn:2008jp} the integral \eqref{Hk} was solved for
$k=0$. In that case the solution immediately follows from the
orthogonality relation \eqref{portho}, resulting in
\begin{align}
{\mathcal H}_{\nu,0}^{(2)}(\alpha,m) = - d_{N-1} 
= \frac{N^{N}}{(N-1)! (1-\alpha)^{N}} \int_0^{(1-\alpha)^2/4\alpha} dt \,
e^{-m^2Nt/(1-\alpha)} \frac{t^{N-1}}{(t+1)^{N+\nu}} \,,
\label{Hk0}
\end{align}
where we substituted the explicit expression for $d_{N-1}$ given in eq.~\eqref{dm}.
To compute the phase of the fermion determinant for arbitrary $N_f$ and $\nu$ we also need the solution of eq.~\eqref{Hk} for positive integer $k$. In this case the second polynomial in the weighted integral is $p_{N-1}^{\nu,k}(z^*;\alpha)$, which involves an orthogonal polynomial  of order $\nu+k$ instead of $\nu$ and an additional power of $z^*$, such that the orthogonality relation \eqref{portho} can no longer be applied directly. 
To solve the integral \eqref{Hk} for arbitrary $k \in \mathbb{N}$ we use the relation 
\begin{align}
p_{\ell}^{\nu,k}(z;\alpha) = z^{2k} p_{\ell}^{\nu+k}(z;\alpha)
&= \sum_{j=0}^{k} \binom{k}{j} \frac{(\ell+\nu+k)!}{(\ell+\nu+k-j)!}
\left(\frac{\alpha-1}{N}\right)^j p_{\ell+k-j}^{\nu}(z;\alpha) 
\label{pbinom}
\end{align}
for any $\nu, \ell, k \in \mathbb{N}$, which expresses the LHS as a sum over orthogonal polynomials of order $\nu$ and degree $\ell,\ldots,\ell+k$. 
The proof of this relation is given in appendix~\ref{App:rel}.  We now substitute this expansion in eq.~\eqref{Hk} to find
\begin{align}
{\mathcal H}_{\nu,k}^{(2)}(\alpha,m) &= 
- \frac{1}{r_{N-1}^\nu(\alpha)} 
\sum_{j=0}^{k} \binom{k}{j} \frac{(N-1+\nu+k)!}{(N-1+\nu+k-j)!}
\left(\frac{\alpha-1}{N}\right)^j \notag\\ 
&\quad\times \sum_{n=0}^{\infty} d_n   
\int_\mathbb{C} d^2z \, w^\nu(z,z^*;\alpha)  
p_n^\nu(z;\alpha) p_{N-1+k-j}^{\nu}(z^*;\alpha) \notag\\
&= - \frac{1}{r_{N-1}^\nu(\alpha)} 
\sum_{j=0}^{k} \binom{k}{j} \frac{(N-1+\nu+k)!}{(N-1+\nu+k-j)!}
\left(\frac{\alpha-1}{N}\right)^j 
d_{N-1+k-j} \: r_{N-1+k-j}^\nu(\alpha)
\,,
\label{Hnuk2b}
\end{align}
where in the last step we have applied the orthogonality relation
\eqref{portho}, after which only the $n=N-1+k-j$ term survives.  Using
the norm \eqref{rknu} we compute the ratio
\begin{align}
\frac{r_{N-1+k-j}^\nu(\alpha)}{r_{N-1}^\nu(\alpha)} 
= \frac{(1+\alpha)^{2(k-j)}(N-1+k-j)!(N-1+k-j+\nu)!}{N^{2(k-j)}(N-1)!(N-1+\nu)!}
\,,
\label{normratio}
\end{align}
and after substituting this and the expression \eqref{dm} for $d$  in
eq.~\eqref{Hnuk2b} we find
\begin{align}
{\mathcal H}_{\nu,k}^{(2)}(\alpha,m)
&= \frac{N^{N-k}}{(1-\alpha)^{N+k} } \frac{(N-1+\nu+k)!}{(N-1)!(N-1+\nu)!}
 \int_0^{(1-\alpha)^2/4\alpha} dt \, e^{-m^2Nt/(1-\alpha)} 
\frac{t^{N-1}}{(t+1)^{N+k+\nu}} 
  \notag\\ & \quad \times
\sum_{j=0}^{k} (-)^j \binom{k}{j} 
(1+\alpha)^{2(k-j)} t^{k-j} 
(1-\alpha)^{2j} (t+1)^j \notag\\
&=   N^{N-k} (4\alpha)^{k+\nu} (1-\alpha)^{N+k} \frac{(N-1+\nu+k)!}{(N-1)!(N-1+\nu)!}\notag\\
&\quad\times\int_0^1 du \,
e^{-m^2 a u} 
\frac{u^{N-1}(u-1)^k}{[(1-\alpha)^2 u +4\alpha]^{N+k+\nu}}  
 \,.
\label{HkN2}
\end{align}
In the last step we have performed the sum over $j$ using the binomial
theorem to obtain $[ 4\alpha t - (1-\alpha)^2 ]^k$.  Also, we have
introduced the variable transformation $u=4\alpha t/(1-\alpha)^2$ to
simplify the integration range.

The sum of ${\mathcal H}_{\nu,k}^{(1)}$ and ${\mathcal H}_{\nu,k}^{(2)}$ of eqs.~\eqref{HkN1} and \eqref{HkN2} gives the exact finite-$N$ result for ${\mathcal H}_{\nu,k}$ needed to compute the average phase of the fermion determinant with eqs.~\eqref{unqphfac2}, \eqref{Z1by1} and \eqref{phappN}. For $\nu <0$ one just needs to replace $\nu$ by $|\nu|$.
Note that the chiral limit ($m \to 0$) of the phase factor \eqref{phappN} has to be taken carefully, as detailed in appendix~\ref{App:chiralN}.

\section{Microscopic limit}
\label{sec:micro}

The results computed from chRMT are universal, i.e., identical to the
corresponding quantities in QCD, in the so-called microscopic regime.
This regime is obtained by defining the rescaled parameters $\hm=2N
m$, $\hm_f = 2N m_f$, $\ha = 2N \alpha$ and the rescaled eigenvalues
$\hz=2N z$ and then taking $N\to\infty$ while keeping the rescaled
quantities fixed.  The conversion of the rescaled random matrix
parameters to the physical parameters is done using the relations
$\hz=z_\text{phys}V\Sigma$, $\hm=m_\text{phys}V\Sigma$ and
$\ha={\hat\mu}^2=\mu_\text{phys}^2F^2V$, where $V$ is the four-volume
and $\Sigma$ and $F$ are low-energy constants of chiral perturbation
theory.
The Gell-Mann--Oakes--Renner relation $m_\pi^2F^2=2m\Sigma$ (for equal
quark masses) can be used to introduce the physical pion mass $m_\pi$
through the combination $\mu_\text{phys}^2/m_\pi^2={\hat\mu}^2/2\hm$.

To take the microscopic limit of the average phase we introduce the corresponding limits of the various objects defined in section~\ref{Sec:RMT} and section~\ref{Sec:Phase}. A detailed derivation of these limits can be found in ref.~\cite[appendix~A]{Bloch:2008cf}.
The microscopic limit (denoted by a subscript $s$) of the orthogonal polynomials \eqref{pknu} is defined as
\begin{align}
  \hp^\nu(\hz;\ha) &= \lim_{N\to\infty}
  \frac{e^N}{(2N)^{\nu+1/2}} p_{N-1}^\nu(\hz/2N;\ha/2N)
  = \sqrt{\pi} e^{-\ha/2} \hz^{-\nu} I_\nu\left(\hz\right) \,. 
  \label{App:limp2}
\end{align}
Accordingly, the microscopic limit of $p_{N-1}^{\nu,k}$, defined in eq.~\eqref{plnuk}, is 
\begin{align}
  \hp^{\nu,k}(\hz;\ha) &= \lim_{N\to\infty}
   \frac{e^N}{(2N)^{\nu-k+1/2}} p_{N-1}^{\nu,k}(\hz/2N;\ha/2N)
  =\sqrt{\pi} e^{-\ha/2} \hz^{-\nu} I_{\nu,k}\left(\hz\right)\,,
\label{limpext}
\end{align}
where we introduced the notation 
\begin{align}
  \label{eq:Inuk}
  I_{\nu,k}(z) = z^k I_{\nu+k}(z)\,. 
\end{align} 
The microscopic limit of the weight function \eqref{wnu} is 
\begin{align}
  \hw^\nu(\hz, \hz^*;\ha) 
  &=\lim_{N\to\infty} (2N)^{2\nu+2} w^\nu(\hz/2N, \hz^*/2N;\ha/2N)
  = |\hz|^{2(\nu+1)} 
  e^{-\frac{\hz^2+\hz^{*2}}{8\ha}}
  K_\nu\left(\frac{|\hz|^2}{4\ha}\right) \,,
\label{App:micw}
\end{align}
and the microscopic limit of the normalization factor \eqref{rknu} is 
\begin{align}
  \hr^\nu(\ha) &= \lim_{N\to\infty} (2N)^2e^{2N}r_{N-1}^\nu(\ha/2N) 
  = 4\pi^2 \ha e^{\ha} \,.
\label{App:micrknu}
\end{align}
We define the microscopic limit of the integral \eqref{HnukN} as
\begin{align}
{\mathcal H}^s_{\nu,k}(\ha,\hm) 
&= \lim_{N\to\infty} e^{-N} (2N)^{\nu+k-1/2}\sqrt{\pi}e^{-\ha/2}\hm^{-\nu} {\mathcal H}_{\nu,k}(\ha/2N,\hm/2N) \,,
\label{miclimit}
\end{align}
where the prefactors were chosen such that it coincides 
with the master integral of ref.~\cite[eq.~(3.21)]{Bloch:2008cf} and remains finite when $N \to \infty$, i.e.,
\begin{align}
{\mathcal H}^s_{\nu,k}(\ha,\hm) &= - \frac{e^{-2\ha}}{4\pi\ha\hm^\nu}
\int_\mathbb{C} \frac{d^2z}{z^2-\hm^2} \frac{|z|^{2(\nu+1)}}{{z^*}^{\nu}} e^{-\frac{z^2+z^{*2}}{8\ha}} K_\nu\left(\frac{|z^2|}{4\ha}\right)  I_{\nu,k}(z^*) \,.
\label{HG}
\end{align}
Here we also used the microscopic limits \eqref{limpext}, \eqref{App:micw} and \eqref{App:micrknu}. Using eqs.~\eqref{limpext} and \eqref{miclimit} we now construct the microscopic limit of the phase factor derived in section~\ref{Sec:Phase}.
For the unquenched case we take the microscopic limit of eq.~\eqref{unqphfac2} and find
\begin{align}
  \phs_{N_f}
  &= \lim_{N\to\infty} \bigl\langle e^{2i\theta} \bigr\rangle_{N_f}
  =  \frac{
 \begin{vmatrix}
    {\mathcal H}^s_{\nu,0}(\ha,\hm) & {\mathcal H}^s_{\nu,1}(\ha,\hm) & \cdots & {\mathcal H}^s_{\nu,\Nf+1}(\ha,\hm)\\[1mm]
    I_{\nu,0}(\hm) & I_{\nu,1}(\hm) & \cdots & I_{\nu,\Nf+1}(\hm)\\[1mm]
    I_{\nu,0}(\hm_1) & I_{\nu,1}(\hm_1) & \cdots & I_{\nu,\Nf+1}(\hm_1)\\
    \vdots & \vdots & \vdots & \vdots \\
    I_{\nu,0}(\hm_\Nf) & I_{\nu,1}(\hm_\Nf) & \cdots & I_{\nu,\Nf+1}(\hm_\Nf)
  \end{vmatrix}
  }
  {\left[\prod_{f=1}^\Nf (\hm_f^2-\hm^2)\right]
    \det\big[I_{\nu,g-1}(\hm_f)\big]_{f,g=1,\ldots,N_f}}\,.
\label{phfacNfflavors_arbmass}
\end{align}
As expected, the dependence on $N$ has dropped out, leaving a finite
microscopic limit for the average phase factor.
Similarly, the microscopic limit of the quenched average phase factor \eqref{Z1by1} is given by
\begin{align}
  \phs_{N_f=0} &= \lim_{N\to\infty} \bigl\langle e^{2i\theta}
  \bigr\rangle_{N_f=0}
  = \begin{vmatrix}
    {\mathcal H}^s_{\nu,0}(\ha,\hm) & {\mathcal H}^s_{\nu,1}(\ha,\hm) \\[1mm]
    I_{\nu,0}(\hm) & I_{\nu,1}(\hm)
  \end{vmatrix} \,.
  \label{qu0}
\end{align}

For the special case in which all dynamical fermions have the same mass $\hm$ as the valence quark, eq.~\eqref{phfacNfflavors_arbmass} becomes
\begin{align}
  \phs_{N_f}
  &= 
  \frac{ \begin{vmatrix}
      \I{0} & \I{1} & \cdots & \I{\Nf+1}\\[1mm]
      I_{\nu,0}(\hm) & I_{\nu,1}(\hm) & \cdots & I_{\nu,\Nf+1}(\hm)\\[1mm]
      I'_{\nu,0}(\hm) & I'_{\nu,1}(\hm) & \cdots & I'_{\nu,\Nf+1}(\hm)\\
      \vdots & \vdots & \vdots & \vdots \\
      I^{(\Nf)}_{\nu,0}(\hm) & I^{(\Nf)}_{\nu,1}(\hm) & \cdots & I^{(\Nf)}_{\nu,\Nf+1}(\hm)
    \end{vmatrix} }
  {(2\hm)^\Nf \Nf!\det\big[I^{(f-1)}_{\nu,g-1}(\hm)\big]_{f,g=1,\ldots,N_f}}
  \,,
  \label{phfacNfflavors}
\end{align}
in analogy to section~\ref{Sec:Phase}. Again, an alternative way to write this result is
\begin{align}
  \phs_{N_f} = \frac{1}{(2\hm)^\Nf \Nf!} \frac{{\mathcal
      W}_{\Nf}(\ha,\hm)}{W_\Nf(0, 1, \ldots, \Nf-1)} \,,
\label{phapp}
\end{align}
where we have defined
\begin{align}
{\mathcal W}_{\Nf}(\ha,\hm) = \sum_{k=0}^{\Nf+1} (-)^k \I{k} W_{\Nf+1}(0,
\ldots, k-
1, k+1, \ldots, \Nf+1)
\label{numerator}
\end{align}
in analogy to \eqref{numeratorN} and $W_{n}(k_1,\ldots,k_n)$ is a
short-hand notation for the Wronskian
\begin{align}
  W_{n}(I_{\nu,k_1}(\hm), \ldots, I_{\nu,k_n}(\hm))=
  \det\big[I^{(i-1)}_{\nu,k_j}(\hm)\big]_{i,j=1,\ldots,n}
  \label{G.4}
\end{align}
appearing in eqs.~\eqref{phapp} and \eqref{numerator}.

To compute the microscopic limit \eqref{phfacNfflavors_arbmass} of the average phase we need to take the microscopic limit \eqref{miclimit} of the complex integral \eqref{HnukN}, which was solved for finite $N$ by eqs.~\eqref{HkN1} and \eqref{HkN2}. 
Taking the microscopic limit \eqref{miclimit} of eqs.~\eqref{HkN1} and
\eqref{HkN2} gives
\begin{align}
  \I{k}
&= \frac{e^{-2\ha-\tfrac{\hm^2}{8\ha}} }{4\ha}
\int_0^{\hm}  du \,  u^{k+1} e^{-\tfrac{u^2}{8\ha}} K_\nu\left(\frac{\hm u}{4\ha}\right) I_{\nu+k}(u)  \notag\\
&\quad + \frac{(4\ha)^{\nu+k} }{2\hm^{\nu}}
\int_0^1 du \, e^{-\tfrac{2\ha}{u}-\tfrac{\hm^2u}{8\ha}} \frac{(u-1)^k}{u^{\nu+k+1}} 
 \,,
\label{Z1totalnew}
\end{align}
where in the first term we have substituted eqs.~\eqref{limpext} and
\eqref{App:micrknu} and in the second term 
we have used the definition 
\begin{equation}
  \lim_{N\to\infty} \left(1-\frac{\ha}{N} \right)^{N} =
  e^{-\ha}
  \label{limexp2}
\end{equation}
of the exponential function and Stirling's formula
\begin{equation}
  \lim_{N\to\infty} \frac{N!}{\sqrt{2\pi N} N^N e^{-N}} = 1
  \,.
\label{limNNN}
\end{equation}
For the special case of $k=0$, eq.~\eqref{Z1totalnew} reproduces the
result which was computed previously in ref.~\cite{Osborn:2008jp} in a
study of the chiral condensate.  However, for the calculation of the
average phase using eq.~\eqref{phfacNfflavors_arbmass} we need the more general result of eq.~\eqref{Z1totalnew} for arbitrary $k \in \mathbb{N}$. The chiral limit of the phase factor \eqref{phapp} is computed in appendix~\ref{app:chiral}.

In ref.~\cite{Bloch:2008cf} we computed the complex integral
\eqref{HG} in a different way, which led to an integral representation that looks quite different from eq.~\eqref{Z1totalnew}. In the next section we will prove the equivalence of the expressions derived in both formulations.

Apart from providing an alternative formula to compute the microscopic
limit of the average phase, the current derivation has the significant additional feature 
that it also gives a calculable and well-behaved expression for
finite-sized matrices, away from the microscopic limit, given by the
sum of eqs.~\eqref{HkN1} and \eqref{HkN2}. This expression will be useful when verifying the analytical RMT results by dynamical RMT simulations in section~\ref{sec:results}.

\newcommand{\expsab}{e^{-\tfrac{a}{s}-b s}}
\newcommand{\expsa}{e^{-\tfrac{a}{s}}}
\newcommand{\expsb}{e^{-bs}}
\newcommand{\cI}{{\mathcal I}}
\newcommand{\J}{{\mathcal J}}

\section{Equivalence of integration results}
\label{sec:equiv}

In ref.~\cite{Bloch:2008cf} the complex integral \eqref{HG} was solved using a completely different formalism, based on the deformation of integration contours and some involved complex analysis, leading to
\begin{align}
  \label{Z1totalold}
  \I{k}
  &= \frac{e^{-2\ha-\tfrac{\hm^2}{8\ha}}}{4\ha} \bigg[
  \int_0^{\hm} du \,
  u^{k+1}
  e^{-\tfrac{u^2}{8\ha}}
  K_\nu\left(\frac{\hm u}{4\ha}\right)
  I_{\nu+k}(u) \\
  &\qquad\qquad\qquad + \int_0^\infty du \, 
  (-)^k u^{k+1} 
  e^{-\tfrac{u^2}{8\ha}}
  I_\nu\left(\frac{\hm u}{4\ha}\right)
  K_{\nu+k}(u)  
  \bigg] \notag\\
  &\quad+ e^{-2\ha-\tfrac{\hm^2}{8\ha}} 
  \frac{(-)^k 2^{\nu-1+k}}{\hm^\nu} 
  \sum_{i,j=0}^{i+j\le\nu-1}
  \frac{(\nu-1-i)!(\nu-1+k-j)!}{(\nu-1-i-j)! i! j!}
  \left(\frac{\hm^{2}}{8\ha}\right)^i \left(2\ha\right)^j
  . \notag
\end{align}
This solution must be equivalent to \eqref{Z1totalnew}, as they are
just two different representations of the same complex integral
\eqref{HG}. Nevertheless, as both expressions look quite different it
is useful to prove their equivalence, without resorting to the
complicated complex analysis used in ref.~\cite{Bloch:2008cf}.  This
proof will also provide a check on both results.

The first integral in eq.~\eqref{Z1totalnew} is identical to the first
integral in eq.~\eqref{Z1totalold}. Therefore it remains to show that
the second integral in eq.~\eqref{Z1totalnew} equals the sum of the
second integral and the additional polynomial in
eq.~\eqref{Z1totalold}.  Setting $a=2\ha$ and $b=\hm^2/8\ha$ to
simplify the notation and canceling some prefactors, we thus need to
show
\begin{align}
a^{\nu+k}
\int_0^1 du \, e^{-\tfrac{a}{u}-bu} \frac{(1-u)^k}{u^{\nu+k+1}} 
  \: \stackrel{?}{=} \: &
e^{-a-b}\bigg[\frac{(ab)^{\nu/2}}{2^k a} 
  \int_0^\infty \!\!du \,   u^{k+1} 
  e^{-\tfrac{u^2}{4a}}
  I_\nu\big(u\sqrt{b/a}\big)
  K_{\nu+k}(u)  
  +S_{\nu,k}\bigg],
 \label{equiv}
\end{align}
where we have defined
\begin{align}
 S_{\nu,k} = \sum_{i,j=0}^{i+j\le\nu-1}
  \frac{(\nu-1-i)!(\nu-1+k-j)!}{(\nu-1-i-j)! i! j!} b^i a^j \,.
\label{Snuk}
\end{align}
For the special case of $\nu=k=0$ this was proven in ref.~\cite{Osborn:2008jp}. Below we will give a general proof for arbitrary $\nu$ and $k$. 

In appendix \ref{Integrel} we prove the integral relation
\begin{align}
  & \int_0^\infty \!\!du \,   u^{k+1} 
  e^{-\tfrac{u^2}{4a}}
  I_\nu\big(u\sqrt{b/a}\big)
  K_{\nu+k}(u) 
\notag\\
&\qquad= 2^k k! 
a \left(\frac{b}{a}\right)^{\nu/2} \!\!\!  e^{a+b} 
\int_0^1 ds \, 
\expsab \, s^{\nu+k-1} L_k^{\nu}\big(-b(1-s)\big) \,.
\label{H2aLk}
\end{align}
Renaming $u$ to $s$ on the LHS of eq.~\eqref{equiv} and substituting \eqref{H2aLk} in the
first term on the RHS, 
proving \eqref{equiv} corresponds to proving the identity
\begin{align}
\J_{\nu,k} &\equiv k! \; b^{\nu} \!\int_0^1 \!ds \, 
\expsab \, s^{\nu+k-1} L_k^{\nu}\big(- b (1-s)\big)
\stackrel{?}{=} a^{\nu+k} \!\int_0^1 \!ds \, 
\expsab \frac{(1-s)^k}{ s^{\nu+k+1}} - e^{-a-b} S_{\nu,k} \,.
\label{arbnu_arbk}
\end{align}
Using the Rodrigues formula for the generalized Laguerre polynomials,
\begin{align}
L_k^{\nu}(x) = \frac{e^x x^{-\nu} }{k!} \frac{d^k}{dx^k} (e^{-x} x^{k+\nu})\,,
\label{RodriguesGeneral}
\end{align}
the LHS of eq.~\eqref{arbnu_arbk} can be written as
\begin{align}
\J_{\nu,k} &=
(-)^k b^{\nu} \int_0^1 ds \, 
\expsa \, 
 \frac{ s^{\nu+k-1} }{ (1-s)^{\nu} }  \frac{d^k}{ds^k} \big[ \expsb (1-s)^{k+\nu} \big]
  \notag\\
& = 
b^{\nu} \int_0^1 ds \, 
\frac{d^k}{ds^k} \left[\expsa \, \frac{ s^{\nu+k-1} }{ (1-s)^{\nu} } \right]  
\expsb (1-s)^{k+\nu} 
  \,,
\label{Inuk}
\end{align}
where in the last step we have performed $k$ successive integrations
by parts.
In appendix~\ref{App:Dnuk} we prove that
\begin{align}
\frac{d^k}{ds^k} \left[\expsa \, \frac{ s^{\nu+k-1} }{ (1-s)^{\nu} } \right]
=  \expsa \sum_{j=0}^k \binom{k}{j} \frac{(\nu)_j a^{k-j} s^{\nu-k+j-1}}{(1-s)^{\nu+j}}\,,
\label{Dnuk}
\end{align}
with the Pochhammer symbol $(\nu)_j$ defined in appendix \ref{Integrel}, 
and substituting this in the integral \eqref{Inuk} we obtain
\begin{align}
\J_{\nu,k} = 
b^{\nu} \int_0^1 ds \, 
 \expsab \sum_{j=0}^k \binom{k}{j} (\nu)_j a^{k-j} s^{\nu-k+j-1} (1-s)^{k-j} 
  \,.
\label{Inuksum}
\end{align}
With this identity we show in appendix~\ref{App:recur} that $\J_{\nu,k}$ satisfies the recurrence relation
\begin{align}
\J_{\nu,k} 
&= - b^{\nu-1} e^{-a-b} (\nu)_k + \sum_{j=0}^k (j+1)_{k-j} 
\left[ a b  \J_{\nu-2,j} + (\nu-1) \J_{\nu-1,j} \right] \,.
\label{conjA}
\end{align}

We now prove the equivalence \eqref{arbnu_arbk} by induction in
$\nu$ using the recurrence relation \eqref{conjA} and the expression
\eqref{Inuksum} for $\J_{\nu,k}$.
For $\nu=0$ only the $j=0$ term contributes in eq.~\eqref{Inuksum}, and we find  
\begin{align}
\J_{0,k} = 
a^{k} \int_0^1 ds \,  \expsab  s^{-k-1} (1-s)^{k} 
\,,
\end{align}
which corresponds to the RHS of eq.~\eqref{arbnu_arbk} since $S_{0,k}=0$. For $\nu=1$ eq.~\eqref{Inuksum} gives
\begin{align}
\J_{1,k} &= 
- \int_0^1 ds \, \bigl[ \expsb \bigr]' \expsa \sum_{j=0}^k
\binom{k}{j} j! \, a^{k-j} \left(\frac1s-1\right)^{k-j} \notag \\
 &= - e^{-a-b} \,  k!  
+ \int_0^1 ds \, \expsab \sum_{j=0}^k \frac{k!}{(k-j)!}
a^{k-j}\left(\frac1s-1\right)^{k-j}
\left[ \frac a{s^2} - \frac{k-j}{s(1-s)} 
\right] \notag\\
&= - e^{-a-b} \,  k!  
+ \int_0^1 ds \, \expsab  \, 
\Biggl[ \sum_{j=0}^k - \sum_{j=1}^k \Biggr]
\frac{k!}{(k-j)!} a^{k-j+1}
  \frac{(1-s)^{k-j}}{s^{k-j+2}} \notag\\
&= - e^{-a-b} \,  k!  + a^{k+1} \int_0^1 ds \, \expsab  \,
\frac{(1-s)^{k}}{s^{k+2}} \,,
\end{align}
where in the second step we have integrated by parts and in the third
step we have observed that the $j=k$ contribution in the last term
vanishes and shifted $j\to j-1$ for this term so that in the last step
only the $j=0$ term of the first sum remains.  For $\nu=1$
eq.~\eqref{Snuk} gives $S_{1,k} = k!$ so that
eq.~\eqref{arbnu_arbk} is satisfied for $\nu=1$.
We now prove that the equivalence holds for $\J_{\nu,k}$ if it is satisfied for $\J_{\nu-1,k}$ and $\J_{\nu-2,k} $. We start from the recurrence relation \eqref{conjA}
and substitute the RHS of eq.~\eqref{arbnu_arbk} for $\nu-1$ and $\nu-2$,
\begin{align}
\J_{\nu,k} 
&=  \sum_{j=0}^k  a^{\nu-1+j} (j+1)_{k-j} 
\int_0^1 ds \, \expsab \frac{(1-s)^j}{ s^{\nu+j}}
\left[   bs +    (\nu-1)  \right] \notag\\
&\quad -  e^{-a-b} \Biggl\{ \sum_{j=0}^k (j+1)_{k-j} \left[ a b S_{\nu-2,j} + (\nu-1) S_{\nu-1,j} \right]
+ b^{\nu-1} (\nu)_k \Biggr\} \,.
\label{eq:jnk1}
\end{align}
Performing an integration by parts on the first term in the integral
yields
\begin{align}
  -\int_0^1 ds \, \bigl[\expsb\bigr]' \expsa \frac{(1-s)^j}{
    s^{\nu+j-1}}
  &=\!\int_0^1 ds \, \expsab \frac{(1-s)^j}{ s^{\nu+j-1}}
  \left[\frac a{s^2}-\frac{j}{1-s}-\frac{\nu+j-1}s\right]
  -e^{-a-b}\delta_{j0}\,,
\label{eq:ints}
\end{align}
where we observed that the surface term is only nonzero for $j=0$.
Let us denote the first term on the RHS of eq.~\eqref{eq:jnk1} by
$\J_{\nu,k}^{(1)}$.  Substituting \eqref{eq:ints} into this term we
obtain
\begin{align}
\J_{\nu,k}^{(1)}
&=  \sum_{j=0}^k a^{\nu-1+j} (j+1)_{k-j}  \int_0^1 ds \, \expsab
\frac{(1-s)^j}{ s^{\nu+j-1}}\left[\frac a{s^2}-\frac{j}{s(1-s)}\right]
-e^{-a-b}a^{\nu-1}k! \notag\\
&=\int_0^1 ds \, \expsab 
\Biggl[\sum_{j=0}^k-\sum_{j=0}^{k-1}\Biggr]a^{\nu+j} (j+1)_{k-j}
\frac{(1-s)^j}{ s^{\nu+j+1}}
-e^{-a-b}a^{\nu-1}k!\notag\\
&=a^{\nu+k}\int_0^1 ds \, \expsab 
\frac{(1-s)^k}{ s^{\nu+k+1}}
-e^{-a-b}a^{\nu-1}k!\,.
\end{align}
In the first line, we observed that the $j=0$ contribution to the
second term in square brackets vanishes, which allowed us to shift
$j\to j+1$ in the corresponding sum in the second line.  In the third
line only the $j=k$ term of the first sum survives.  Putting
everything together, we have
\begin{align}
\J_{\nu,k} 
&=   a^{\nu+k}  \int_0^1 ds \, \expsab \frac{(1-s)^k}{ s^{\nu+k+1}}\notag\\ 
&\quad-  e^{-a-b} \Biggl\{ \sum_{j=0}^k (j+1)_{k-j} \left[ a b
  S_{\nu-2,j} + (\nu-1) S_{\nu-1,j} \right]  
+ b^{\nu-1} (\nu)_k +  a^{\nu-1} k! \Biggr\}\,.
\end{align}
To reproduce the RHS of eq.~\eqref{arbnu_arbk} it remains to show that
\begin{align}
\sum_{j=0}^k (j+1)_{k-j} \left[ a b S_{\nu-2,j} + (\nu-1)
  S_{\nu-1,j} \right] + b^{\nu-1} (\nu)_k + a^{\nu-1}k!  
\stackrel{?}{=} S_{\nu,k} \,,
\label{Sigma_equiv}
\end{align}
which we relegate to appendix~\ref{app:sigma}.  This completes the
proof of the equivalence \eqref{arbnu_arbk} for arbitrary $\nu$ and
$k$.

\section{Dynamical random matrix simulations with complex weights}
\label{sec:results}

\subsection{Computing averages using reweighting}

In ref.~\cite{Bloch:2008cf} the analytical chRMT results for the
microscopic limit of the quenched phase factor were thoroughly
verified using quenched random matrix simulations. In this section we
verify the unquenched analytical results using dynamical random matrix
simulations. In the dynamical case the finite-$N$ corrections to the
microscopic results are quite significant.  Increasing the matrix size
in the simulations to be close to the microscopic limit would require
too much computational power, especially since dynamical simulations
are already intrinsically expensive.  Instead, we chose to use
small-sized matrices and compare the results for the average phase
factor with the finite-$N$ predictions of eq.~\eqref{phappN}, where
the master integral is given by the sum of eqs.~\eqref{HkN1} and
\eqref{HkN2}.  Even then, the random matrix simulations at nonzero
chemical potential are problematic since the partition function
contains a complex weight.  Both the real and imaginary parts of the
weight function can become negative, and as a consequence the fermion
determinant cannot be included in the probability distribution of a
Markov Chain Monte Carlo (MCMC) simulation, i.e., we are confronted
with the sign problem.

To circumvent this problem we will perform the importance sampling
using an auxiliary, nonnegative, weight function, and reweight the
result appropriately so that we average over the correct target
ensemble. This reweighting leads to an \textit{overlap} problem, when
the configurations contributing most to the partition functions in
both ensembles do not coincide.  Note that this overlap problem always
occurs in reweighting, even if the weight in the target ensemble is
nonnegative. This is, for example, also an issue at zero chemical
potential when using dynamical fermions with a heavier mass in the
auxiliary ensemble than in the target ensemble. In the presence of a
chemical potential, the overlap problem is amplified by the sign
problem. Even if the auxiliary distribution has a good overlap with
the target ensemble, i.e., the relevant configurations are sampled
appropriately, the sign problem can ruin the reweighting
procedure. This will happen due to large cancellations of
contributions with opposite sign in the partition function of the
target ensemble, which will generate large statistical errors.

Below we briefly describe the principle of reweighting.\footnote{For
  related approaches see, e.g., \cite{Anagnostopoulos:2001yb,Ambjorn:2002pz,Fodor:2007vv,Ejiri:2007ga,Anagnostopoulos:2010ux}.}  Using the
partition function \eqref{partfun} we want to compute the unquenched
expectation value
\begin{align}
\langle{\mathcal O}\rangle_{\Nf} = \frac{1}{Z_\nu^{\Nf}}\int d\phi_1 d\phi_2 w(\phi_1) w(\phi_2)  {\mathcal D}[\phi_1,\phi_2;\mu;\{m_f\}]  {\mathcal O}[\phi_1,\phi_2]
\end{align}
with dynamical determinant
\begin{align}
{\mathcal D}[\phi_1,\phi_2;\mu;\{m_f\}] = \prod_{f=1}^{N_f} \det (D(\mu)+m_f) \equiv R e^{i\Theta} \,.
\label{fermdet}
\end{align}
Here, $R$ is the (nonnegative) magnitude and $e^{i\Theta}$ is the
phase of $\mathcal D$.  In the case of interest for this study, the
observable will be ${\mathcal O}=e^{2i\theta}$.  To set up the reweighting
formalism we introduce the weighted (or ensemble) average
\begin{align}
\langle {f} \rangle_{w} = \frac{\int dx \; w(x) \, f(x)}{\int dx \; w(x)}
\label{wavg}
\end{align}
of $f$ with respect to $w$, with normalization $\langle 1 \rangle_w = 1$. From this definition we see that an ensemble average can be computed from an auxiliary ensemble using the reweighting relation
\begin{align}
\langle g \rangle_{w f} = \frac{\langle f g \rangle_w}{ \langle f \rangle_w } \,.
\label{fgw}
\end{align}
This feature is useful to study ensembles with weight functions that
cannot be sampled efficiently, or where the weights are not positive
definite such that they cannot be used as probability distributions in
importance sampling.  The actual simulation constructs a Markov chain
for an auxiliary ensemble, after which the expectation value of the
observable in the target ensemble is computed by reweighting the
observable and the partition function as given in eq.~\eqref{fgw}. To
keep the statistical error of the reweighted observable within
reasonable limits, the overlap between both ensembles should be large,
i.e., the bulk of relevant configurations in both ensembles should
coincide.

In this study we compare the results obtained with three different
reweighting schemes \cite{Bloch:2009th}.  The ensembles and the
corresponding reweightings are as follows.
\begin{description}\labelwidth9mm\itemindent0mm
\item[R1]
Quenched simulation with full reweighting: We perform a standard quenched simulation of random matrices through direct sampling of the Gaussian weights \eqref{rmtdis}  for the real and imaginary parts of the elements of $\phi_1$ and $\phi_2$, as  described in appendix~E of ref.~\cite{Bloch:2008cf}, and reweight with the dynamical fermion determinant \eqref{fermdet}. Using eq.~\eqref{fgw} the unquenched average is computed from
\begin{align}
\langle{\mathcal O}\rangle_{\Nf} = \frac{\langle R e^{i\Theta} {\mathcal O}\rangle_{\Nf=0}}{\langle R e^{i\Theta}\rangle_{\Nf=0}} \,,
\label{fullrew}
\end{align}
where the quenched averages in the numerator and denominator are
evaluated as averages over the Monte Carlo sample.

The omission of the fermion determinant in the importance sampling
will cause an overlap problem, and the reweighting factors will
strongly fluctuate between configurations. These fluctuations will
increase as the chemical potential grows and the sign problem becomes
more tangible. In this reweighting scheme the generation of the
matrices in the Markov chain is cheap, but we need a very long chain
to get an acceptable accuracy.

\item[R2] Phase quenched simulation using Metropolis with
  partial reweighting: In this Metro\-polis algorithm the matrix
  probability distribution consists of the product of the Gaussian
  weights \eqref{rmtdis} and the magnitude $R$ of the dynamical
  determinant. The measurement is then reweighted by the phase factor
  of the dynamical determinant,
\begin{align}
\langle{\mathcal O}\rangle_{\Nf} = \frac{\langle e^{i\Theta} {\mathcal O}\rangle_{R}}{\langle e^{i\Theta}\rangle_{R}}
 \,.
\label{Iphq}
\end{align}
Including information about the determinant in the sampling probability should improve the overlap between the generated configurations and the significant configurations in the unquenched ensemble.

\item[R3] Sign quenched simulation using Metropolis with minimal
  reweighting: As the unquenched partition function \eqref{partfun} is
  real, the contributions from the determinant \eqref{fermdet} to the
  partition function only come from $R\cos\Theta$. The imaginary
  contributions cancel between Hermitian conjugate matrices as these
  have the same Gaussian probability \eqref{rmtdis}. To minimize the
  variance of the reweighting factors and optimize the overlap it is
  therefore natural to create an auxiliary ensemble using the weight
  $R |\cos\Theta|$ and absorb the remaining sign of the determinant in
  the reweighting factor,
\begin{align}
\langle{\mathcal O}\rangle_{\Nf} = 
\frac{\langle (\sign\cos\Theta)(1+i\tan\Theta)\;{\mathcal O}\rangle_{R|\cos\Theta|}}{\langle \sign\cos\Theta\rangle_{R|\cos\Theta|}}
 \,,
\label{algA4}
\end{align}
where the denominator is explicitly real because of the symmetry
mentioned above.  Using the absolute value of the weights as auxiliary
distribution allows one to include as much information as possible
about the determinant in the MCMC weights.  Note that the average
over the $i\tan\Theta$ term in the numerator does not vanish in this
case because the observable $\mathcal O=e^{2 i\theta}$ also has an imaginary
component. We expect this reweighting scheme to be somewhat more
effective for real than for complex observables.

After performing simulations using this reweighting, we
realized\footnote{We thank Philippe de Forcrand for bringing this
  earlier study to our attention.} that this idea of minimal
reweighting had been discussed earlier \cite{deForcrand:2002pa} in a
study of the reweighting factor in lattice QCD simulations at nonzero
$\mu$. In that study is was shown, based on the central limit theorem,
that reweighting by $R|\cos\Theta|$ indeed minimizes the fluctuations
in the reweighting factor.  In ref.~\cite{deForcrand:2002pa} this
reweighting scheme was not implemented since it would have been too
expensive in lattice QCD.  Here, we put it to a practical test on the
random matrix model.  (Note that scheme R3 was also rediscovered in
\cite{Hsu:2010zza}.)

Note that for a complex observable one can develop more sophisticated
reweighting algorithms using two Markov chains for the numerator, one
for the real part of the weights and another for the imaginary
part. The complication is that the chains have to be normalized
properly with respect to each other, which introduces additional
overhead. This will not be investigated further here.

\end{description}

Another partial reweighting scheme uses the $\mu$-quenched ensemble,
where the matrices are sampled using a dynamical MCMC algorithm at
zero chemical potential, and the contributions to the averages are
reweighted by the ratio of the determinants at chemical potential
$\mu$ and at $\mu=0$.  For small $\mu$, this scheme is close to scheme
R2, and for larger $\mu$ it is no better than scheme R1 but much more
expensive.  Therefore it will not be studied here.

We briefly discuss the main reweighting features of the three
schemes. Clearly, the overlap problem is best handled by the sign
quenched scheme, which will sample the most significant configurations
of the partition function. The biggest overlap problem will be
encountered by the quenched scheme, as the fermion determinant is
completely ignored in the auxiliary weight function. Nevertheless, the
construction of the Markov chain is very cheap in the quenched scheme,
so that the overlap problem can be partially alleviated by the
generation of many more configurations, which are all uncorrelated by
construction. Both the sign quenched and phase quenched schemes use a
Metropolis algorithm. Here the autocorrelation time has to be taken
into account to select the independent configurations, which very much
shortens the effective size of the Markov chain. Fortunately, the much
smaller fluctuations allow us to reach a high accuracy with much fewer
uncorrelated configurations than in the quenched scheme. Although the
sign quenched scheme seems superior to the phase quenched one in the
case of random matrix simulations, the latter can be more easily
implemented in realistic theories like QCD. A phase quenched
determinant for $\Nf=2$ can be implemented by simulating a quark and a
conjugate quark, whereas a sign quenched determinant does not seem to
have a physical equivalent which could be implemented efficiently.

Even though the severity of the overlap problem is different for the
three schemes, there is no reason why the sign problem would be
improved upon in any of the schemes. Indeed, even in the sign quenched
case where we sample the most significant configurations, the positive
and negative contributions will balance each other more and more when
the chemical potential becomes large, and the sign problem will
remain. This will be confirmed in the numerical experiments discussed
in the next section.

In the numerical implementation, for each matrix generated in the
importance sampling we also consider its Hermitian conjugate matrix,
in accordance with the symmetries of the partition function. This
ensures that the sample average of the phase factor is explicitly
real, which somewhat simplifies the implementation of the algorithm
and the computation of the statistical error on the final result.

\subsection{Numerical results}

We performed dynamical random matrix simulations at nonzero chemical
potential using the three reweighting schemes described in the
previous section. For each measurement we generated 1,000,000
matrices with sizes ranging from $N=2$ to 16.  In the quenched
ensemble the generated matrices are uncorrelated, but for the sign and
phased quenched ensembles, which are sampled using a Metropolis
algorithm, successive matrices in the Markov chain are
correlated. This autocorrelation effectively reduces the number of
independent configurations and is taken into account appropriately
when computing the statistical errors on the measurements.  In
addition, the statistical errors take into account the correlations
between numerator and denominator in eq.~\eqref{fgw}.  Details of the
calculation of the statistical errors are given in
appendix~\ref{App:err}.

\begin{figure}[t]
  \centerline{\includegraphics[width=0.49\textwidth]{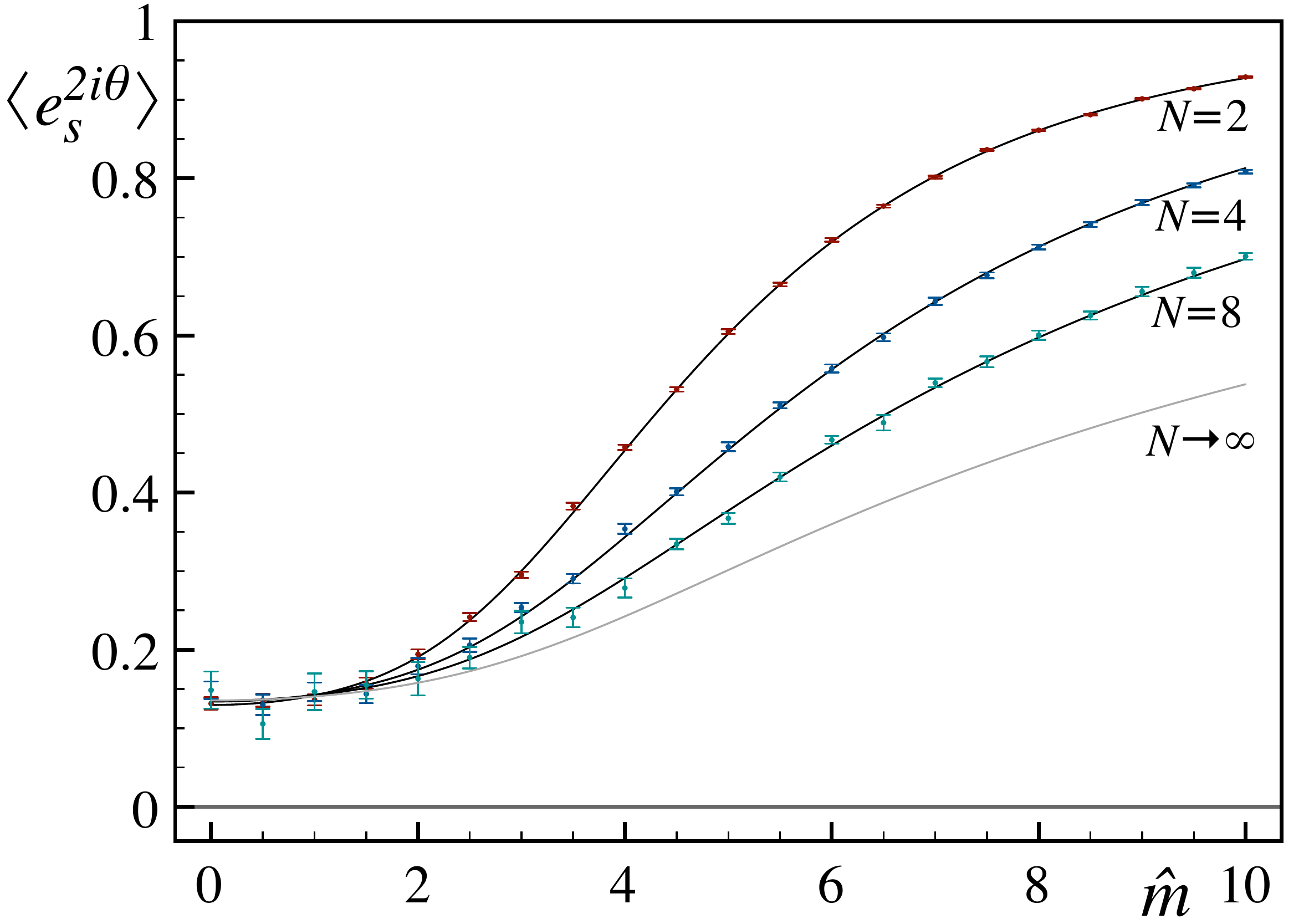}}
  \caption{Average phase factor of the fermion determinant for $\Nf=2$
    as a function of the fermion mass $\hm$ for $\ha=1.0$ and $\nu=1$
    for matrix sizes $N=2$, $4$, and $8$. The simulation results,
    computed with sign quenched reweighting, are given by the data
    points, while the solid lines show the corresponding analytical
    results of eqs.~\eqref{HkN1} and \eqref{HkN2}. We also show the
    microscopic limit ($N\to\infty$) of eq.~\eqref{Z1totalnew}.}
  \label{Fig:unq-m-dep}
\end{figure}

In figure~\ref{Fig:unq-m-dep} we verify the mass dependence of the
phase factor for $N=2$, $4$, and $8$ for $\Nf=2$ and $\nu=1$. The
simulation results agree very well with the analytical predictions of
eq.~\eqref{phappN}. Both the $\hm$-dependence and the dependence on
the matrix size $N$ of the matrices is reproduced. The different
curves show how the large-$N$ limit is approached. For larger masses
the microscopic limit is not yet reached for matrix sizes up to $N=8$,
and the analytical results for finite $N$, away from the microscopic
limit, are essential to explain the numerical results.

\begin{figure}[t]
  \centerline{\includegraphics[width=0.49\textwidth]{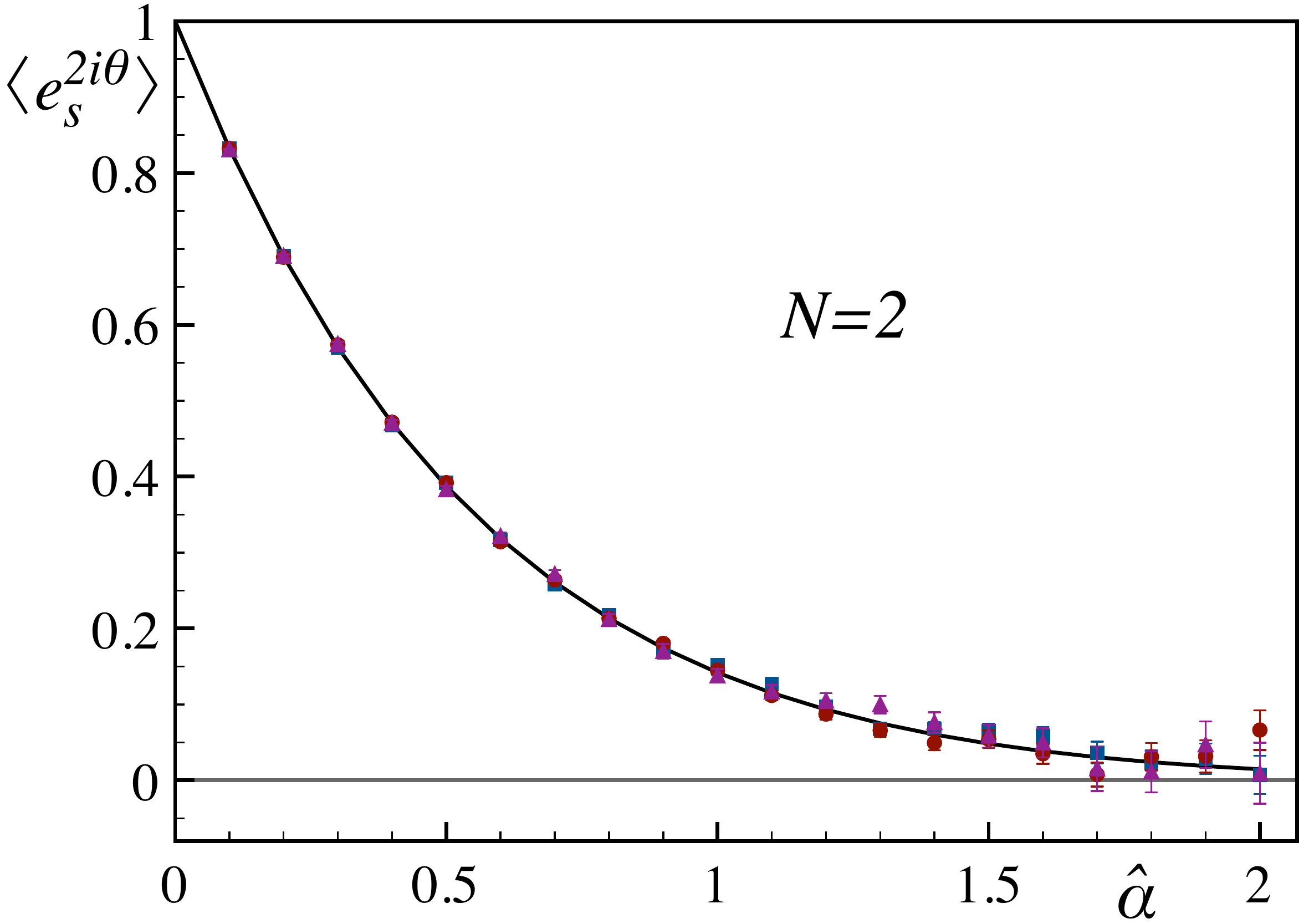}
  \includegraphics[width=0.49\textwidth]{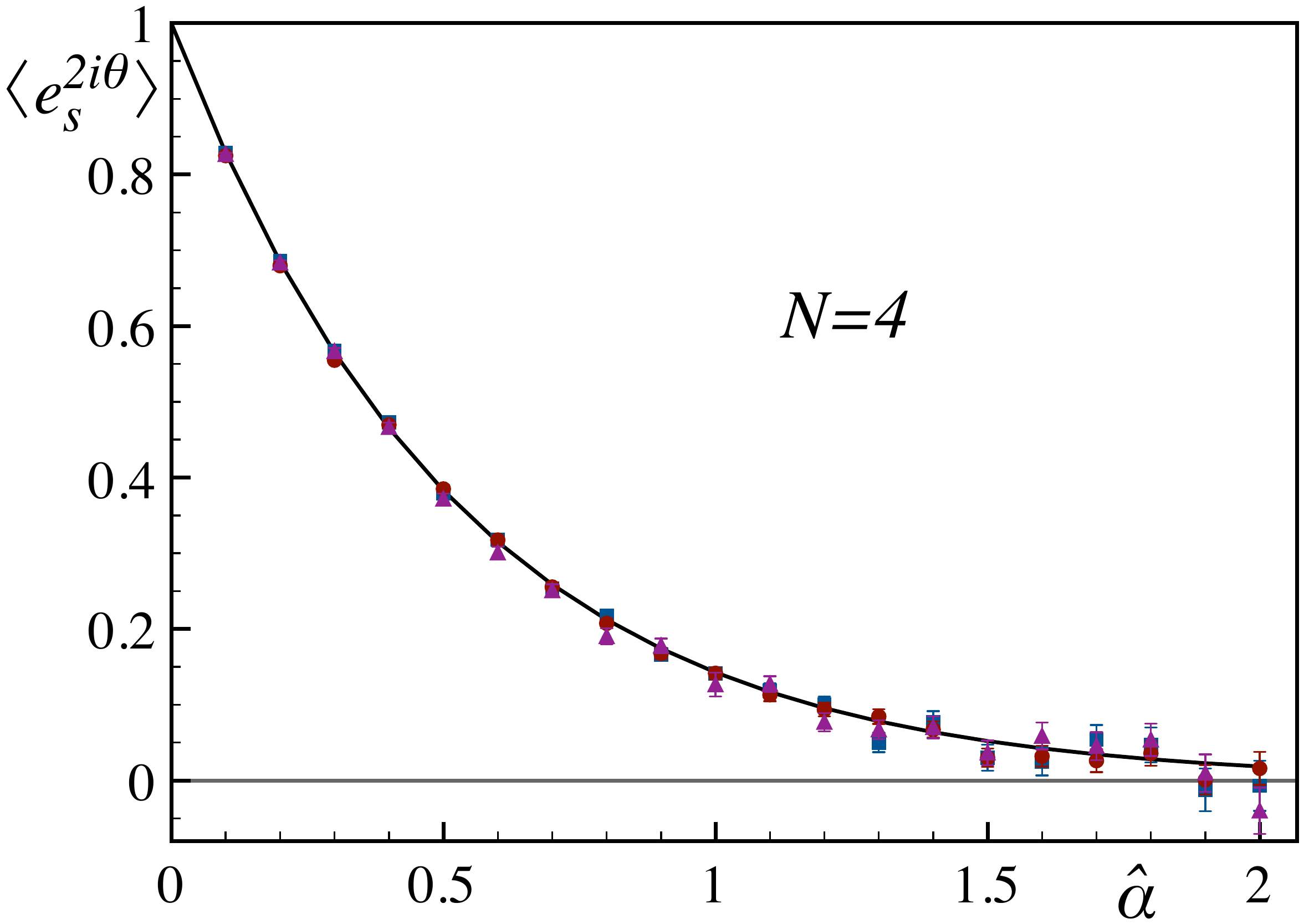}}
  \centerline{\includegraphics[width=0.49\textwidth]{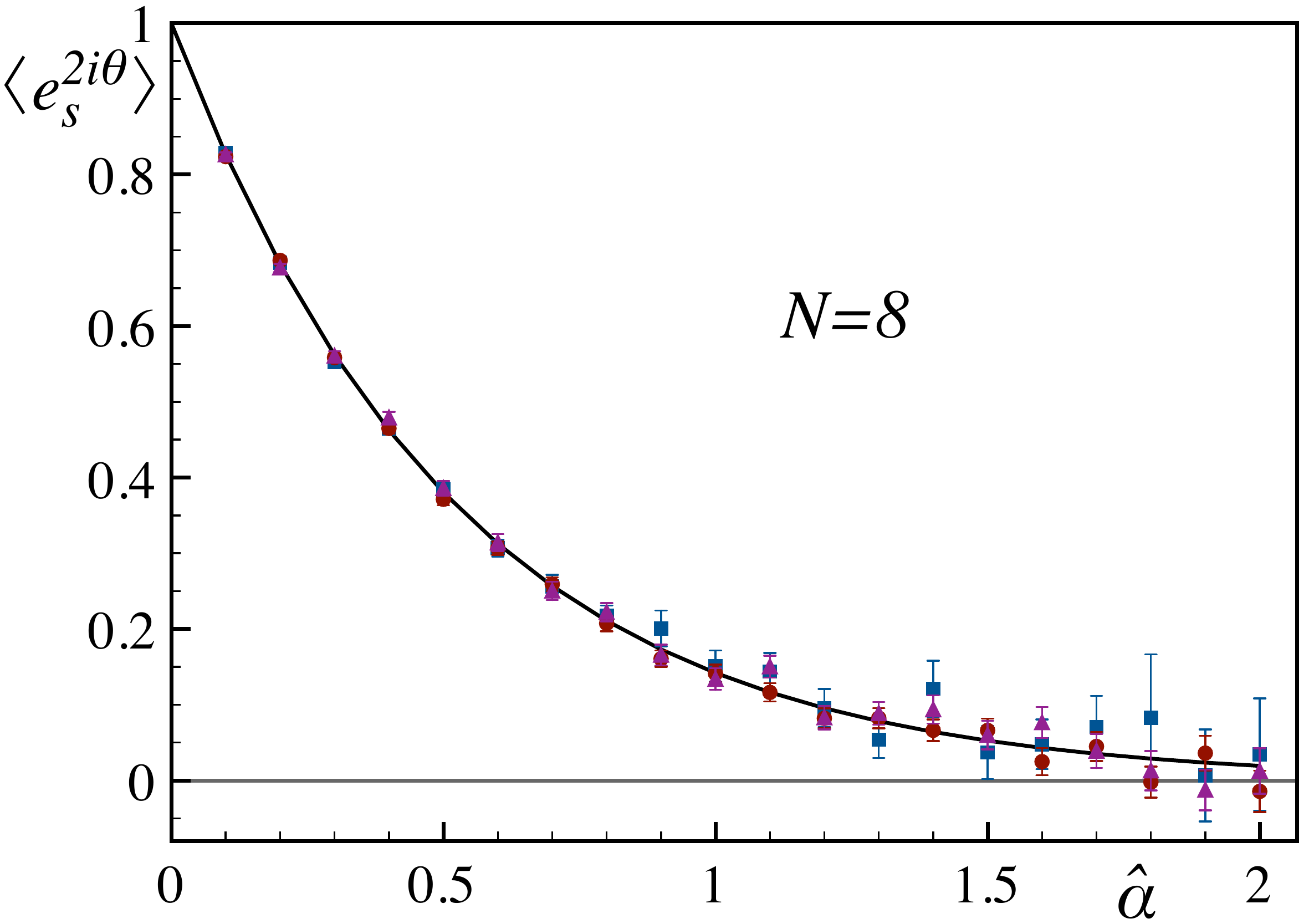}
  \includegraphics[width=0.49\textwidth]{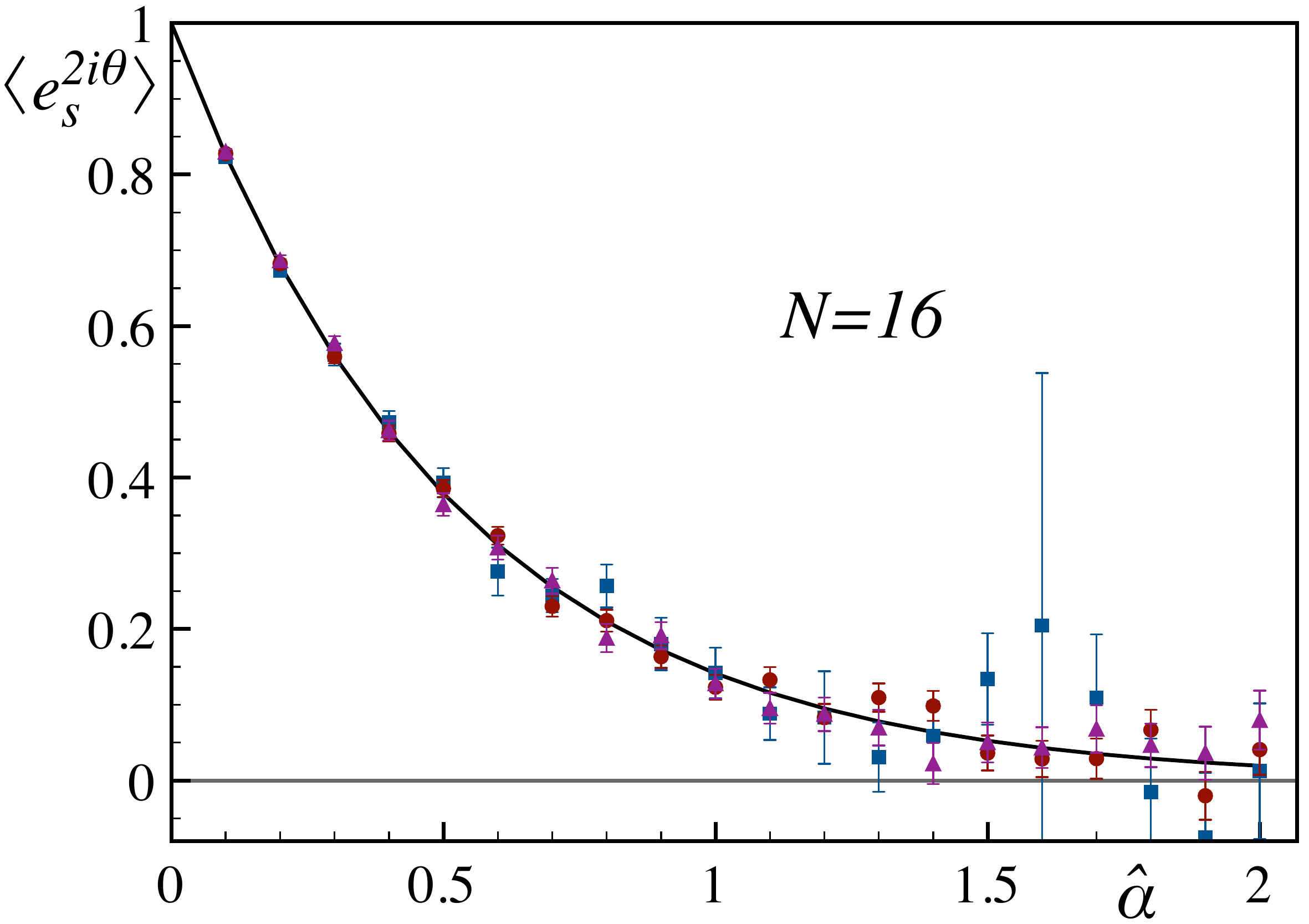}}
  \caption{Average phase factor of the fermion determinant for $\Nf=2$
    as a function of the chemical potential $\ha$ for $\hm=1.0$ and
    $\nu=1$ for matrix sizes $N=2$, $4$, $8$, and $16$ using three
    different reweighting schemes. The blue squares were computed from
    a quenched ensemble with full reweighting, the red circles from a
    phase quenched simulation with partial reweighting, and the purple
    triangles from a sign quenched simulation with minimal
    reweighting. }
  \label{Fig:unq-alg}
\end{figure} 

Figure~\ref{Fig:unq-alg} illustrates how the phase factor changes as a
function of the chemical potential for matrix sizes $N=2$, $4$, $8$,
and $16$. The results again agree very well with the analytical
predictions.  To probe the overlap and sign problems, which manifest
themselves in the accuracy of the measurements, the simulations were
performed using the three reweighting algorithms discussed above. For
small chemical potential the accuracy of all three algorithms is
satisfying, reflecting the fact that neither the overlap nor the sign
problem is significant. However, the phase and sign quenched
algorithms achieve this accuracy with much fewer independent
configurations than the quenched algorithm, as a consequence of the
better overlap between target and auxiliary ensemble. The shorter
trajectories would be an advantage if the measurement itself were
expensive. With increasing chemical potential the error on the
measurements increases as the sign problem sets in (for $\ph \lesssim
0.2$). The sign problem equally affects the different reweighting
methods. As expected, the sign problem gets worse for larger $N$. In
the Metropolis algorithm this deterioration for increasing $N$ is
related to an increasing autocorrelation time.  Even for the sign
quenched algorithm, where the overlap problem is minimized, the sign
problem remains as the matrices are reweighted by $+1$ or $-1$, which
leads to important cancellations and large statistical errors for
large $\ha$.

\section{Conclusions}
\label{Sec:concl}

Dynamical lattice simulations of QCD at nonzero baryon density are
hindered by the sign problem caused by the complex fermion
determinant. To investigate this problem it is helpful to employ the
equivalence between the spectral properties of the Dirac operator in
the $\epsilon$-regime of QCD and chiral random matrix theory, which
also holds at nonzero chemical potential.  As the average phase factor
of the fermion determinant is an important observable in the study of
the sign problem, we have computed it in the framework of chiral
random matrix theory.

In ref.~\cite{Bloch:2008cf} we derived an analytical formula for the
average phase factor in the microscopic limit of quenched and
unquenched chiral random matrix theory for general topology. In the
current paper we presented an alternative derivation, leading to a
different integral representation of this microscopic limit, and
showed that both formulations are equivalent. In contrast to our
previous work, the new formulation also gives calculable expressions
for finite-sized matrices, away from the microscopic limit.

The analytical predictions of the finite-$N$ formula were verified in
dynamical RMT simulations, where reweighting techniques were used to
compute averages in an ensemble with a complex weight. Very good
agreement was found. At large chemical potential the statistical
errors grow, signifying the emergence of the sign problem.

\acknowledgments{ This work was supported in part by the DFG
  collaborative research center SFB/TR-55. We would like to thank
  J. Verbaarschot and Ph. de Forcrand for useful suggestions.  }

\appendix

\section{Relation between orthogonal polynomials}
\label{App:rel}

In this appendix we prove the identity \eqref{pbinom} by induction.
For $k=0$ \eqref{pbinom} trivially holds for any $\nu, \ell \in \mathbb{N}$. For integer $k>0$ the recurrence relation \eqref{pkrecur} yields
\begin{align}
p_{\ell}^{\nu,k}(z;\alpha) = p_{\ell+1}^{\nu,k-1}(z;\alpha) - (\ell+\nu+k) \left(\frac{1-\alpha}{N}\right) p_{\ell}^{\nu,k-1}(z;\alpha) \,.
\end{align}
Assuming that \eqref{pbinom} holds for $p_{\ell}^{\nu,k-1}$ and any $\nu, \ell \in \mathbb{N}$, we can substitute it twice in the previous equation, for $p_{\ell+1}^{\nu,k-1}$ and $p_{\ell}^{\nu,k-1}$, to find
\begin{align}
p_{\ell}^{\nu,k}(z;\alpha) &= 
\sum_{j=0}^{k-1} \binom{k-1}{j} \frac{(\ell+\nu+k)!}{(\ell+\nu+k-j)!}
\left(\frac{\alpha-1}{N}\right)^j p_{\ell+k-j}^{\nu}(z;\alpha) \notag\\
& \quad + \sum_{j=0}^{k-1} \binom{k-1}{j} \frac{(\ell+\nu+k)!}{(\ell+\nu+k-1-j)!}
\left(\frac{\alpha-1}{N}\right)^{j+1} p_{\ell-1+k-j}^{\nu}(z;\alpha) \notag\\
&= 
p_{\ell+k}^{\nu}(z;\alpha) 
+ \frac{(\ell+\nu+k)!}{(\ell+\nu)!}
\left(\frac{\alpha-1}{N}\right)^{k} p_{\ell}^{\nu}(z;\alpha) \notag\\
&\quad+ \sum_{j=1}^{k-1} \left[ \binom{k-1}{j} + \binom{k-1}{j-1} \right] \frac{(\ell+\nu+k)!}{(\ell+\nu+k-j)!}
\left(\frac{\alpha-1}{N}\right)^j p_{\ell+k-j}^{\nu}(z;\alpha) \,, 
\end{align}
where we have first shifted the index $j$ by one in the second sum and
then gathered the overlapping terms in both sums.  As the binomial
coefficients satisfy
\begin{align}
\binom{k-1}{j} + \binom{k-1}{j-1} = \frac{(k-1)!}{j! (k-1-j)!} + \frac{(k-1)!}{(j-1)! (k-j)!} 
= \frac{k!}{j! (k-j)!} = \binom{k}{j} 
\end{align}
we find
\begin{align}
p_{\ell}^{\nu,k}(z;\alpha) &= 
\sum_{j=0}^{k}  \binom{k}{j} \frac{(\ell+\nu+k)!}{(\ell+\nu+k-j)!}
\left(\frac{\alpha-1}{N}\right)^j p_{\ell+k-j}^{\nu}(z;\alpha) 
\label{pbinomk}
\,,
\end{align}
which shows that the identity \eqref{pbinom} holds for arbitrary integer $k\ge 0$.

\section{Chiral limit}

\subsection[Finite $N$]{\boldmath Finite $N$}
\label{App:chiralN}

The chiral limit of the average phase \eqref{phappN} has to be taken carefully such that the mass factors in numerator and denominator are canceled properly.
For this we need to compute the limit $m \to 0$ 
of the Wronskian \eqref{G.4N}, which contains derivatives of the function $p_\ell^{\nu,k}$ defined
in eq.~\eqref{plnuk}.  For small argument $m$ the leading-order
term of this polynomial is
\begin{align}
  p_\ell^{\nu,k}(m;\alpha) \sim \left(\frac{1-\alpha}{N}\right)^{\ell}
  \frac{(\ell+\nu+k)!}{(\nu+k)!} m^{2k} \,,
  \label{Iorigin2}
\end{align}
with $p$-th derivative
\begin{align}
  \partial_m^p\,p_\ell^{\nu,k}(m;\alpha) \sim
  \left(\frac{1-\alpha}{N}\right)^{\ell}\frac{(\ell+\nu+k)!}{(\nu+k)!} 
  \frac{(2k)!}{(2k-p)!} m^{2k-p} \,. 
  \label{InukpLO}
\end{align}
Substituting these expressions in the Wronskian \eqref{G.4N} and using
properties of the determinant gives the leading-order result
\begin{align}
W^\ell_n(k_1,\ldots,k_n) \sim  \left(\frac{1-\alpha}{N}\right)^{\ell n} 
m^{2\sum_i k_i-n(n-1)/2}
\prod_{i=1}^n \frac{(\ell+\nu+k_i)!}{(\nu+k_i)!}
2^{n(n-1)/2} \Delta_n(k_1, \ldots, k_n) \,, 
\label{WchiralN}
\end{align}
where $\Delta_n(k_1, \ldots, k_n)$ is a Vandermonde determinant.  From
this we compute the chiral limit of the denominator in
eq.~\eqref{phappN},
\begin{align}
W^{N}_{N_f}(0, 1,\ldots,\Nf-1) \sim \left(\frac{1-\alpha}{N}\right)^{N N_f}
(2m)^{\Nf(\Nf-1)/2}
\prod_{i=0}^{\Nf-1}\frac{(N+\nu+i)!}{ (\nu+i) !} 
\prod_{\ell=1}^{\Nf-1}
\ell ! \,, 
\label{denomchiralN}
\end{align}
where we used the identities
\begin{align}
  \label{eq:vand}
  \Delta_\Nf(0,1,\ldots,\Nf-1) = \prod_{\ell=1}^{\Nf-1} \ell!
  \quad\text{and}\quad
  \sum_{i=0}^{\Nf-1} i = \frac12\Nf(\Nf-1) \,.
\end{align}
From eq.~\eqref{WchiralN} it is easy to see that in the limit $m \to
0$ only the Wronskian corresponding to the $k=\Nf+1$ term in
eq.~\eqref{numeratorN} contributes to leading order, while all
other terms are of higher order. The average phase \eqref{phappN}
can therefore be written as
\begin{align}
\ph_{m=0}
&= \lim_{m\to 0} \frac{ (-)^{\Nf+1} }{(2m)^\Nf \Nf!} 
  \frac{ W^{N-1}_{\Nf+1}(0, 1, \ldots, \Nf)}{W^N_\Nf(0, 1, \ldots, \Nf-1)}
   \,{\mathcal H}_{\nu,\Nf+1}(\alpha,m) \notag\\
&= (-)^{\Nf+1}
\left(\frac{1-\alpha}{N}\right)^{N-1-N_f}  
\frac{ (N-1+\nu)! }{ (\nu+\Nf) ! }
\,{\mathcal H}_{\nu,\Nf+1}(\alpha,m=0)     
\,,
\label{phfacchiralN}
\end{align}
where in the last step we have substituted \eqref{denomchiralN}.
It now remains to take $m=0$ in ${\mathcal H}$ computed using the
integrals \eqref{HkN1} and \eqref{HkN2}. The first integral is zero as
the integration range vanishes, and hence the average phase simplifies to
\begin{align}
\ph_{m=0}
&= 
\left[\frac{(1-\alpha)^2}{4\alpha}\right]^N 
\frac{(N+\nu+\Nf)!}{(N-1)! (\nu+\Nf) !} \notag\\
&\quad\times\int_0^1  du \,
u^{N-1} \left(1 - u \right)^{\Nf+1} 
\left[1+\frac{(1-\alpha)^2 u}{4\alpha}\right]^{-N-\Nf-\nu-1}\notag\\
&= 
\left[\frac{(1-\alpha)^2}{4\alpha}\right]^N 
\frac{(\Nf+1)!(N+\Nf+\nu)!}{ (\Nf+\nu)! (N+\Nf+1)!} \notag\\
&\quad\times{}_2F_1\left(N+\Nf+\nu+1,N;N+\Nf+2;-\frac{(1-\alpha)^2}{4\alpha}\right)
\notag\\
&= \left(\frac{1-\alpha}{1+\alpha}\right)^{2N} 
\left[\frac{4\alpha}{(1+\alpha)^2}\right]^{\nu-1}
\frac{ (N_f+1)! (N+N_f+\nu)!}{(N_f+\nu)!(N+N_f+1)!} \notag\\
&\quad\times{}_2F_1\left(1-\nu, \Nf+2; N+\Nf+2; -\frac{(1-\alpha)^2}{4\alpha}
 \right) ,
\label{eq:chN}
\end{align}
where in the second step we have used the integral representation of
the hypergeometric function \cite[eq.~(15.3.1)]{Abram:1964} and in the
last step we have applied the transformation
${}_2F_1(a,b;c;z)=(1-z)^{c-a-b}{}_2F_1(c-a,c-b;c;z)$
\cite[eq.~(15.3.3)]{Abram:1964}.  For $\nu\ge 1$ the hypergeometric
function is a polynomial of degree $\nu-1$, see eq.~\eqref{eq:sum}
below, and for $\nu=1$ the result further simplifies to the
$\Nf$-independent expression
\begin{align}
\ph_{\nu=1, m=0} 
&= \left(\frac{1-\alpha}{1+\alpha}\right)^{2N}
\,.
\label{nu=1N}
\end{align}

\subsection{Microscopic limit}
\label{app:chiral}

We now take the $N\to\infty$ limit of the results in
appendix~\ref{App:chiralN} with $\ha=2N\alpha$ fixed.  In this limit
we can replace $(1+\alpha)^2$ in the second factor and $(1-\alpha)^2$ in
the fourth factor of \eqref{eq:chN} by 1.  We also have
\begin{align}
  \lim_{N\to\infty}\left(\frac{1-\alpha}{1+\alpha}\right)^{2N}
  =e^{-2\ha}\,.
\end{align}
For $\nu=0$ we use the relation
\begin{align}
  \lim_{N\to\infty}{}_2F_1\left(1,n;N;-\frac Nx\right)
  &=\lim_{N\to\infty}(N-1)\int_0^1dt\frac{(1-t)^{N-2}}{(1+Nt/x)^n}
  =x^ne^x\int_x^\infty dz\frac{e^{-z}}{z^n}\notag\\
  &=x^ne^x\Gamma(-n+1,x)\,,
  \label{eq:lim}
\end{align}
where in the first step we have used the integral representation of
${}_2F_1$ \cite[eq.~(15.3.1)]{Abram:1964}, in the second step we have
substituted $z=x+Nt$ and taken the $N\to\infty$ limit, and in the
third step we have used the integral representation of the incomplete
gamma function \cite[eq.~(6.5.3)]{Abram:1964}.  With \eqref{eq:lim} we
obtain from \eqref{eq:chN}
\begin{align}
  \phs_{\nu=0,\hm=0} &= e^{-2\ha}\left[\lim_{N\to\infty}
    \frac{N}{2\ha}\frac{N_f+1}{N+N_f+1}\right]
  (2\ha)^{N_f+2}e^{2\ha}\Gamma(-N_f-1,2\ha)\notag\\
  &=(N_f+1)(2\ha)^{N_f+1}\Gamma(-N_f-1,2\ha)\,.
  \label{eq:phfacchiralnu=0}
\end{align}
For $\nu=1$ we obtain immediately from \eqref{nu=1N}
\begin{align}
  \phs_{\nu=1,\hm=0} = e^{-2\ha}\,.
  \label{eq:phfacchiralnu=1}
\end{align}
For $\nu\ge2$ the first argument of the hypergeometric function in
\eqref{eq:chN} is a negative integer, in which case we have
\cite[eq.~(15.4.1)]{Abram:1964}
\begin{align}
  {}_2F_1(-m,b;c;z)=\sum_{n=0}^m\frac{(-m)_n(b)_n}{(c)_n}\frac{z^n}{n!}
  \label{eq:sum}
\end{align}
and obtain
\begin{align}
  \phs_{\hm=0}&=e^{-2\ha}\lim_{N\to\infty}
  \frac{(N_f+1)!(N+N_f+\nu)!}{(N_f+\nu)!(N+N_f+1)!}
  \sum_{n=0}^{\nu-1} \frac{(-)^n}{n!}\frac{(1-\nu)_n(N_f+2)_n}{(N+N_f+2)_n}
  \left(\frac{2\ha}N\right)^{\nu-1-n}\notag\\
  &=e^{-2\ha}\sum_{n=0}^{\nu-1}\frac{(\nu-1)!(N_f+n+1)!}
  {(N_f+\nu)!n!(\nu-1-n)!}(2\ha)^{\nu-1-n}
  \underset{=1}{\underbrace{\lim_{N\to\infty}
      \frac{(N+N_f+\nu)!}{(N+N_f+n+1)!N^{\nu-1-n}}}}\notag\\
  &=e^{-2\ha}\sum_{j=0}^{\nu-1}\binom{\nu-1}j
  \frac{(N_f+\nu-j)!}{(N_f+\nu)!}(2\ha)^j\,,
  \label{eq:phfacchiralnu}
\end{align}
where in the last line we have substituted $j=\nu-1-n$.

Eqs.~\eqref{eq:phfacchiralnu=0}, \eqref{eq:phfacchiralnu=1} and
\eqref{eq:phfacchiralnu} agree with the results previously computed
in ref.~\cite{Bloch:2008cf}.

\section{Proof of integral relation (\protect\ref{H2aLk})}
\label{Integrel}

Below we prove the integral relation \eqref{H2aLk}. In the following, the LHS of 
eq.~\eqref{H2aLk} will be denoted by $\cI_{\nu,k}$, i.e.,
\begin{align}
\cI_{\nu,k} = \int_0^\infty \!\!du \, u^{k+1} 
  e^{-\tfrac{u^2}{4a}}
  I_\nu\big(u\sqrt{b/a}\big)
  K_{\nu+k}(u) \,.
\label{H2aLk1}
\end{align}
Using the integral representation \cite[eq.~(8.432.6)]{Grad:1980}
\begin{align}
K_\nu(u) = \frac{1}{2}\left(\frac{u}{2}\right)^\nu \int_0^\infty \frac{dt}{t^{\nu+1}} e^{-t-\tfrac{u^2}{4t}} \qquad \text{if } \re(u^2) > 0 \,,
\label{intKnu}
\end{align}
which is valid for arbitrary $\nu$, we obtain
\begin{align}
\label{H2a}
\cI_{\nu,k}
&= 
\frac{1}{2^{\nu+k+1}} 
\int_0^\infty \frac{dt}{t^{\nu+k+1}}  e^{-t}
\int_{0}^\infty du \, u^{\nu+2k+1}  e^{-\tfrac{t+a}{4a t}u^2}
I_\nu\big(u\sqrt{b/a}\big)  \,.
\end{align}
The integral over $u$ is given in \cite[eq.~(6.631.1)]{Grad:1980} in
terms of a confluent hypergeometric function,
\begin{align}
  &\int_{0}^\infty du \, u^{\nu+2k+1} e^{-\tfrac{t+a}{4a t}u^2} 
  I_\nu \big(u\sqrt{b/a}\big)\notag\\
  &\qquad=\frac{\Gamma(\nu+k+1)}{2^{\nu+1}\Gamma(\nu+1)}
  \left(\frac ba\right)^{\nu/2}
  \left(\frac{4a t}{t+a}\right)^{\nu+k+1}
  {}_1F_1\left(\nu+k+1;\nu+1;\frac{bt}{t+a}\right).
\end{align}
Using the Kummer transformation \cite[eq.~(13.1.27)]{Abram:1964}
\begin{align}
_1F_1(a;b;z) = e^z {}_1F_1(b-a;b;-z)
\end{align}
we rewrite \eqref{H2a} as
\begin{align}
\cI_{\nu,k}
&=
\frac{\Gamma(\nu+k+1)}{\Gamma(\nu+1)}
  2^{k} a^{\nu+k+1} \left(\frac ba\right)^{\nu/2}
\int_0^\infty \frac{dt}{(t+a)^{\nu+k+1}}  e^{-t+\tfrac{bt}{t+a}}
  {}_1F_1\left(-k;\nu+1;-\frac{bt}{t+a}\right) \notag\\
&= 
\frac{\Gamma(\nu+k+1)}{\Gamma(\nu+1)} 2^{k} a \left(\frac ba\right)^{\nu/2} e^{a+b}
\int_0^1 ds \, \expsab \, s^{\nu+k-1} 
 {}_1F_1\left(-k;\nu+1;-b(1-s)\right) , 
\end{align}
where in the last step we have introduced the variable transformation
$s=a/(t+a)$.
Note that ${}_1F_1\left(-k;\nu+1;z\right) = k! L_k^{\nu}(z)/(\nu+1)_k$
with the Pochhammer symbol $(a)_n = a(a+1)\ldots(a+n-1)$ and $(a)_0=1$
so that the integral finally becomes
\begin{align}
\cI_{\nu,k}
&= 2^{k} k! a \left(\frac ba\right)^{\nu/2} \!\! e^{a+b} 
\int_0^1 ds \, 
\expsab \, s^{\nu+k-1} L_k^{\nu}\big(-b(1-s)\big) \,.
\label{H2aLk2}
\end{align}
This proves the integral relation \eqref{H2aLk}.

\section{Proof of identity (\protect\ref{Dnuk})}
\label{App:Dnuk}

In this appendix we prove the identity
\begin{align}
D_{\nu,k} \equiv 
\frac{d^k}{ds^k} \left[\expsa \, \frac{ s^{\nu+k-1} }{ (1-s)^{\nu} } \right]
=  \expsa \sum_{j=0}^k \binom{k}{j} \frac{(\nu)_j a^{k-j} s^{\nu-k+j-1}}{(1-s)^{\nu+j}}
\label{App-Eq:Dnuk}
\end{align}
by induction. 
For $k=0$ the identity is trivially satisfied for any $\nu$, with
\begin{align}
D_{\nu,0} =\expsa \, \frac{ s^{\nu-1} }{ (1-s)^{\nu} } \,.
\end{align}
For $k \geq 1$ we perform one derivative in $D_{\nu,k}$ explicitly and obtain
\begin{align}
D_{\nu,k}  
&= \frac{d^{k-1}}{ds^{k-1}} \left[
a \expsa \, \frac{ s^{\nu+k-3} }{ (1-s)^{\nu} }
+ (\nu+k-1) \expsa \, \frac{ s^{\nu+k-2} }{ (1-s)^{\nu} }
+ \nu \expsa \, \frac{ s^{\nu+k-1} }{ (1-s)^{\nu+1} } 
\right]
\notag\\
&= \frac{d^{k-1}}{ds^{k-1}} \left[
a \expsa \, \frac{ s^{\nu+k-3} }{ (1-s)^{\nu-1} }
+ (a+\nu+k-1) \expsa \, \frac{ s^{\nu+k-2} }{ (1-s)^{\nu} }
+ \nu \expsa \, \frac{ s^{\nu+k-1} }{ (1-s)^{\nu+1} } 
\right]
\notag\\
&= a D_{\nu-1,k-1} + (a+\nu+k-1) D_{\nu,k-1} + \nu D_{\nu+1,k-1} \,.
\end{align}
Assuming that eq.~\eqref{App-Eq:Dnuk} holds for $k-1$ we thus have
\allowdisplaybreaks[4]
\begin{align}
D_{\nu,k}
&= \expsa \sum_{j=0}^{k-1} \binom{k-1}{j} 
\biggl[\frac{(\nu-1)_j a^{k-j} s^{\nu-k+j-1}}{(1-s)^{\nu+j-1}}
+ (a+\nu+k-1)  \frac{(\nu)_j a^{k-j-1} s^{\nu-k+j}}{(1-s)^{\nu+j}} \notag\\
&\qquad\qquad\qquad\qquad\quad
+ \frac{(\nu)_{j+1} a^{k-j-1} s^{\nu-k+j+1}}{(1-s)^{\nu+j+1}} \biggr] \notag\\
&= \expsa \Biggl\{  \sum_{j=0}^{k-1} \binom{k-1}{j} 
 a^{k-j} \left[ \frac{(\nu-1)_j s^{\nu-k+j-1}}{(1-s)^{\nu+j-1}}
+ \frac{(\nu)_j s^{\nu-k+j}}{(1-s)^{\nu+j}} \right]  \notag \\
&\qquad\qquad + \sum_{j=0}^{k-1} \binom{k-1}{j} a^{k-j-1} \left[ (\nu+k-1)
  \frac{(\nu)_j  s^{\nu-k+j}}{(1-s)^{\nu+j}} 
+ \frac{(\nu)_{j+1} s^{\nu-k+j+1}}{(1-s)^{\nu+j+1}} \right] \Biggr\} \notag\\
&= \expsa  \Biggl\{ \sum_{j=0}^{k-1} \binom{k-1}{j} 
a^{k-j} \left[ \frac{(\nu-1)_j s^{\nu-k+j-1}}{(1-s)^{\nu+j-1}}
+ \frac{(\nu)_j s^{\nu-k+j}}{(1-s)^{\nu+j}} \right] \notag\\
&\qquad\qquad + \sum_{j=1}^{k} \binom{k-1}{j-1} a^{k-j} \left[ (\nu+k-1)
  \frac{(\nu)_{j-1}  s^{\nu-k+j-1}}{(1-s)^{\nu+j-1}} 
+ \frac{(\nu)_{j} s^{\nu-k+j}}{(1-s)^{\nu+j}} \right] \Biggr\} \notag\\
&= \expsa  \Biggl\{
a^{k} \frac{s^{\nu-k-1}}{(1-s)^{\nu}}
+ \sum_{j=1}^{k-1} a^{k-j} \biggl[
\binom{k-1}{j} \left( \frac{(\nu-1)_j s^{\nu-k+j-1}}{(1-s)^{\nu+j-1}}
+ \frac{(\nu)_j s^{\nu-k+j}}{(1-s)^{\nu+j}} \right)  \notag\\
&\qquad\qquad  +  \binom{k-1}{j-1} \left( (\nu+k-1)  \frac{(\nu)_{j-1}
    s^{\nu-k+j-1}}{(1-s)^{\nu+j-1}} 
+ \frac{(\nu)_{j} s^{\nu-k+j}}{(1-s)^{\nu+j}} \right) \biggr]
+ \frac{(\nu)_{k} s^{\nu-1}}{(1-s)^{\nu+k}} 
 \Biggr\}\,,
\end{align}
where in the second step we have gathered the terms with equal powers
of $a$, in the third step we have shifted the index of the second sum,
and in the last step we have extracted the $j=0$ and $j=k$ terms.  The
binomial coefficients satisfy
\begin{align}
\binom{k-1}{j-1} + \binom{k-1}{j} = \binom{k}{j}  \quad \text{and} \quad 
\binom{k}{j} = \frac{k}{j} \binom{k-1}{j-1} \,,
\end{align}
and the Pochhammer symbols obey the identity
\begin{align}
(\nu)_j = (\nu-1)_j \frac{\nu-1+j}{\nu-1} = (\nu-1)_j + j (\nu)_{j-1} \,.
\end{align}
Using these identities we find
\begin{align}
D_{\nu,k}
&= \expsa  \Biggl[ 
a^{k} \frac{s^{\nu-k-1}}{(1-s)^{\nu}}
+ \sum_{j=1}^{k-1} \binom{k}{j} a^{k-j}
 \frac{(\nu)_j s^{\nu-k+j-1}}{(1-s)^{\nu+j}}
+ \frac{(\nu)_{k} s^{\nu-1}}{(1-s)^{\nu+k}} 
 \Biggr] .
\end{align}
The first and the last term can be reabsorbed in the sum, which yields
the identity \eqref{App-Eq:Dnuk}.

\section{\boldmath Recurrence relation for $\J_{\nu,k}$}
\label{App:recur}

\begin{figure}[t]
\centerline{\includegraphics[width=0.5\textwidth]{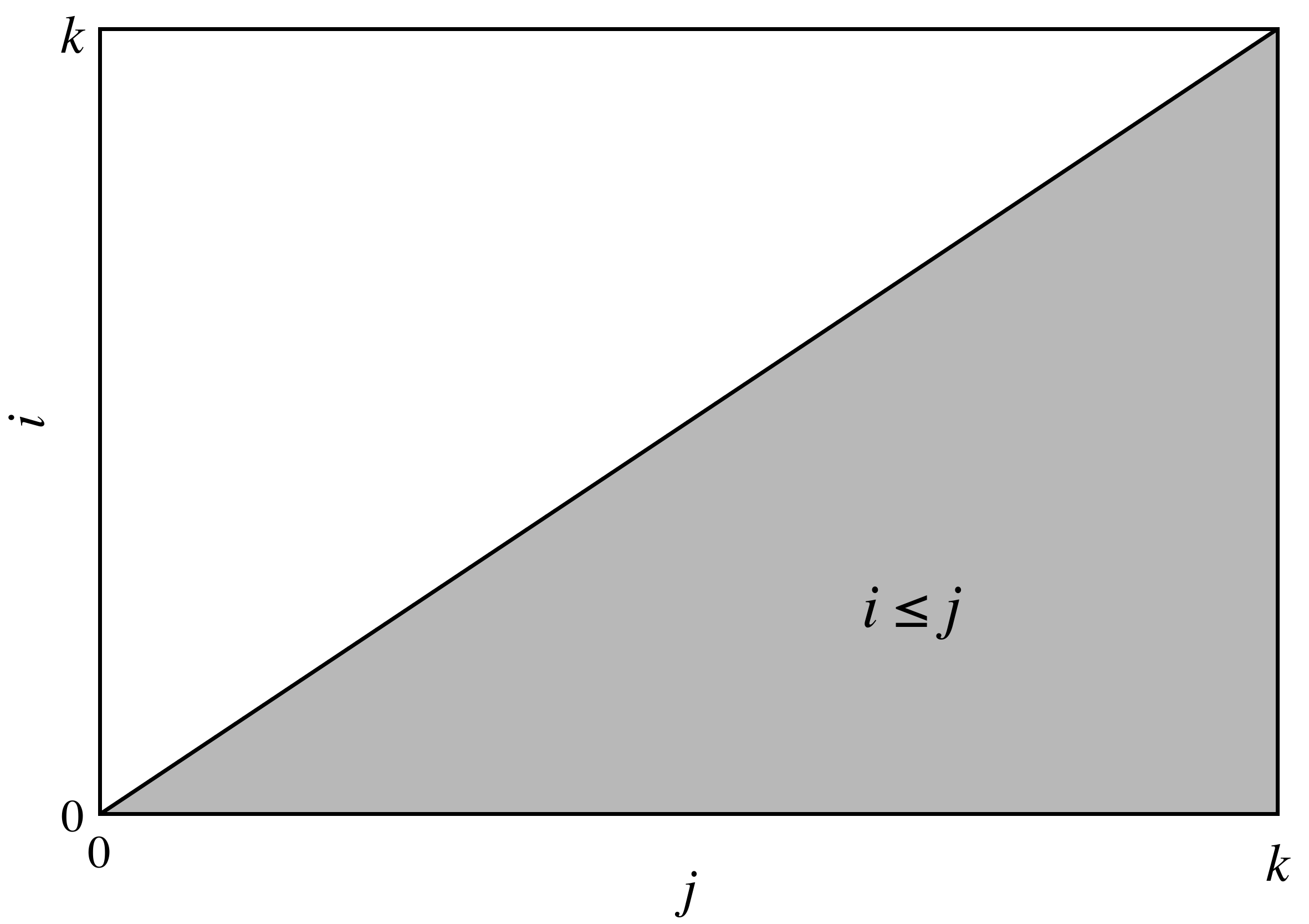}}
\caption{The range of the indices in the double sums of eq.~\protect\eqref{InukA2} is given by the shaded area.}
\label{Fig:indices}
\end{figure}

In this appendix we show that the integral $\J_{\nu,k}$, defined on
the LHS of eq.~\eqref{arbnu_arbk}, satisfies the recurrence relation
\begin{align}
\J_{\nu,k} 
&= - b^{\nu-1} e^{-a-b} (\nu)_k + \sum_{j=0}^k (j+1)_{k-j} 
\left[ a b  \J_{\nu-2,j} + (\nu-1) \J_{\nu-1,j} \right] .
\label{App:conjA}
\end{align}
To prove this relation we substitute eq.~\eqref{Inuksum} for
$\J_{\nu,k}$ and rewrite the integral as
\begin{align}
\J_{\nu,k} &= 
- b^{\nu-1} \int_0^1 ds \, 
 \bigl[\expsb\bigr]' \expsa \sum_{j=0}^k \binom{k}{j} (\nu)_j a^{k-j} s^{\nu-k+j-1} (1-s)^{k-j}\notag\\
&= - b^{\nu-1} e^{-a-b} (\nu)_k  
+ b^{\nu-1} \int_0^1 ds \, 
 \expsb \biggl[ \expsa \sum_{j=0}^k \binom{k}{j} (\nu)_j a^{k-j}
 s^{\nu-k+j-1} (1-s)^{k-j} \biggr]' \notag\\
&= - b^{\nu-1} e^{-a-b} (\nu)_k  + b^{\nu-1} \int_0^1 ds \, 
 \expsab \sum_{j=0}^k \binom{k}{j} (\nu)_j a^{k-j} s^{\nu-k+j-3} (1-s)^{k-j-1}
 \notag\\
 &\quad  \times\left[ a (1-s) +  (\nu-k+j-1)  s (1-s) - (k-j) s^2 \right] ,
\label{eq:sumj}
\end{align}
where in the second step we have integrated by parts (noting that only
the $j=k$ term contributes to the surface term) and in the last step
we have performed the derivatives.  We now manipulate the sum over $j$
in eq.~\eqref{eq:sumj}, which we denote by $\mathcal S$.  We split the
coefficient $(\nu-k+j-1)$ of the second term in $\nu-1$ and $-(k-j)$
and merge this second part with the third term to obtain
\begin{align}
\mathcal S
&=  \sum_{j=0}^k \binom{k}{j} (\nu)_j a^{k-j} s^{\nu-k+j-3}
(1-s)^{k-j-1} \bigl[  a(1-s)+(\nu-1)s(1-s)-(k-j)s \bigr].
\end{align}
Since for the third term the $j=k$ contribution to the sum vanishes we
shift (for the third term only) the upper limit of the sum to $k-1$
and then $j\to j-1$, resulting in
\begin{align}  
  \mathcal S&= \sum_{j=0}^k \binom{k}{j} a^{k-j} s^{\nu-k+j-3} (1-s)^{k-j}
\bigl[(\nu)_ja+(\nu-1)_{j+1}s\bigr]\notag\\
 &\quad - \sum_{j=1}^k \binom{k}{j-1} (k-j+1) (\nu)_{j-1} a^{k-j+1} s^{\nu-k+j-3} (1-s)^{k-j}  \notag\\
&= \sum_{j=0}^k \binom{k}{j} a^{k-j} s^{\nu-k+j-3} (1-s)^{k-j} 
\bigl[(\nu-1)_ja+(\nu-1)_{j+1}s\bigr],
\label{InukA}
\end{align}
where in the last step we 
used the identity
\begin{align}
\binom{k}{j} (\nu)_j - \binom{k}{j-1} (k-j+1) (\nu)_{j-1} 
&= \frac{k!(\nu)_j}{j! (k-j)!} - \frac{k!(\nu)_{j-1}}{(j-1)!(k-j)!}
\notag\\
&= \frac{k!(\nu)_{j-1}}{j! (k-j)!} \left[ (\nu\!+\!j\!-\!1) - j \right]
= \binom{k}{j} (\nu-1)_{j} \,.
\end{align}
The Pochhammer symbols satisfy
\begin{align}
(\nu-1)_j = (\nu-2)_j \frac{\nu-2+j}{\nu-2} = (\nu-2)_j + j (\nu-1)_{j-1} \,,
\end{align}
and repeating this argument $j$ times we find
\begin{align}
(\nu-1)_j = \sum_{i=0}^j (j-i+1)_i (\nu-2)_{j-i}\,.
\label{poch1}
\end{align}
Similarly, 
\begin{align}
(\nu-1)_{j+1} = (\nu-1)_j (\nu-1+j) = (\nu-1)(\nu-1)_j +j (\nu-1)_j \,,
\end{align}
which after $j$ repetitions yields
\begin{align}
(\nu-1)_{j+1} = (\nu-1) \sum_{i=0}^j (j-i+1)_i (\nu-1)_{j-i}\,.
\label{poch2}
\end{align}
Using the identity  
\begin{align}
j \binom{k}{j} = \frac{k!}{(j-1)!(k-j)!} = k \binom{k-1}{j-1}
\end{align}
repeatedly, we find from eq.~\eqref{poch1} 
\begin{align}
\binom{k}{j} (\nu-1)_j 
&= \sum_{i=0}^j \binom{k}{j} (j\!-\!i\!+\!1)_i (\nu-2)_{j-i} 
= \sum_{i=0}^j (k\!-\!i\!+\!1)_i \binom{k-i}{j-i} (\nu-2)_{j-i}
\label{poch3}
\intertext{and from eq.~\eqref{poch2}}
\binom{k}{j} (\nu-1)_{j+1} 
&= (\nu-1) \sum_{i=0}^j \binom{k}{j}(j-i+1)_i (\nu-1)_{j-i} \notag\\
&= (\nu-1) \sum_{i=0}^j (k-i+1)_i \binom{k-i}{j-i} (\nu-1)_{j-i} .
\label{poch4}
\end{align}
We now substitute the identities \eqref{poch3} and \eqref{poch4} in eq.~\eqref{InukA} and find
\begin{align}
\mathcal S
&= \sum_{j=0}^k a^{k-j} s^{\nu-k+j-3} (1-s)^{k-j}
\sum_{i=0}^j (k-i+1)_i \binom{k-i}{j-i} \bigl[ (\nu-2)_{j-i} a
+ (\nu-1)  (\nu-1)_{j-i}  s \bigr] \notag\\
&= \sum_{i=0}^k  (k-i+1)_i  \sum_{j=i}^k  \binom{k-i}{j-i}  a^{k-j}
s^{\nu-k+j-3} (1-s)^{k-j}   
 [(\nu-2)_{j-i}a + (\nu-1) (\nu-1)_{j-i}  s] \notag\\
&= \sum_{i=0}^k  (k-i+1)_i \sum_{j=0}^{k-i}  \binom{k-i}{j}  a^{k-j-i}
s^{\nu-k+j+i-3} (1-s)^{k-j-i} \bigl[
 (\nu\!-\!2)_{j}a + (\nu\!-\!1) (\nu\!-\!1)_{j}  s \bigr] ,
\label{InukA2}
\end{align}
where in the second step we have interchanged the sums over $i$ and
$j$ (see the shaded triangle in figure~\ref{Fig:indices}) and in the
last step we have shifted the index $j$ by $i$.
Looking back at the eq.~\eqref{Inuksum} we see that
$\J_{\nu,k}$ of eq.~\eqref{eq:sumj} can be written as
\begin{align}
\J_{\nu,k} 
&= - b^{\nu-1} e^{-a-b} (\nu)_k  + b^{\nu-1} \int_0^1 ds \, 
 \expsab S \notag\\
&= - b^{\nu-1} e^{-a-b} (\nu)_k
+ \sum_{i=0}^{k} (k-i+1)_i \left[ a b \J_{\nu-2, k-i}
 + (\nu-1) \J_{\nu-1,k-i} 
\right] ,
\end{align}
which, after introducing $j=k-i$, proves the relation \eqref{App:conjA}.

\section{Proof of identity (\protect\ref{Sigma_equiv})}
\label{app:sigma}

With the definitions
\begin{align}
  \label{eq:T}
  T_{\nu,n} &= a b S_{\nu-2,n} + (\nu-1)  S_{\nu-1,n}\\
  \Sigma_{\nu,k} &= \sum_{n=0}^k (n+1)_{k-n} T_{\nu,n}
  \label{Sigma}
\end{align}
we need to show that
\begin{align}
  \Sigma_{\nu,k} + b^{\nu-1} (\nu)_k + a^{\nu-1}k!  = S_{\nu,k} \,.
\end{align}
We substitute the definition~\eqref{Snuk} of $S_{\nu,k}$ in
\eqref{eq:T} and perform some manipulations that are explained at the
end of the string of equations,
\begin{align}
T_{\nu,n} &= 
\sum_{i=0}^{\nu-3} \sum_{j=0}^{\nu-3-i}
\frac{(\nu-3-i)!(\nu-3+n-j)!}{(\nu-3-i-j)! i! j!} b^{i+1} a^{j+1}\notag\\
&\quad + \sum_{i=0}^{\nu-2}\sum_{j=0}^{\nu-2-i}
  \frac{(\nu-1)(\nu-2-i)!(\nu-2+n-j)!}{(\nu-2-i-j)! i! j!} b^i a^j 
  \notag\\
&= \sum_{i=1}^{\nu-2} \sum_{j=0}^{\nu-2-i}
\frac{(\nu-2-i)!(\nu-3+n-j)!}{(\nu-2-i-j)! (i-1)! j!} b^{i} a^{j+1}\notag\\
&\quad + \sum_{i=0}^{\nu-2}\sum_{j=0}^{\nu-2-i}
  \frac{(\nu-1)(\nu-2-i)!(\nu-2+n-j)!}{(\nu-2-i-j)! i! j!} b^i a^j 
  \notag\\
&= \sum_{j=0}^{\nu-2} \frac{(\nu-1)!(\nu-2+n-j)!}{(\nu-2-j)! j!} a^j 
+ \sum_{i=1}^{\nu-2} \Biggl[ 
\sum_{j=0}^{\nu-2-i}
\frac{(\nu-2-i)!(\nu-3+n-j)!}{(\nu-2-i-j)! (i-1)! j!} b^{i} a^{j+1} \notag\\
&\quad + \sum_{j=0}^{\nu-2-i}
  \frac{(\nu-1)(\nu-2-i)!(\nu-2+n-j)!}{(\nu-2-i-j)! i! j!} b^i a^j 
 \Biggr] \notag\\
&= \sum_{j=0}^{\nu-2} \frac{(\nu-1)!(\nu-2+n-j)!}{(\nu-2-j)! j!} a^j
 + \sum_{i=1}^{\nu-2} \Biggl[ 
\sum_{j=1}^{\nu-1-i}
\frac{(\nu-2-i)!(\nu-2+n-j)!}{(\nu-1-i-j)! (i-1)! (j-1)!} b^{i} a^{j} \notag\\
&\quad + \sum_{j=0}^{\nu-2-i}
  \frac{(\nu-1)(\nu-2-i)!(\nu-2+n-j)!}{(\nu-2-i-j)! i! j!} b^i a^j
 \Biggr] \notag\\
&=\sum_{j=0}^{\nu-2} \frac{(\nu-1)!(\nu-2+n-j)!}{(\nu-2-j)! j!} a^j 
+ \sum_{i=1}^{\nu-2} 
\Biggl[\frac{(n-1+i)!}{ (i-1)!} b^{i} a^{\nu-1-i} 
+\frac{(\nu-1)(\nu-2+n)!}{ i! } b^i \notag\\
&\quad  + \sum_{j=1}^{\nu-2-i} 
\bigl( i j +   (\nu-1)(\nu-1-i-j) \bigr)
 \frac{(\nu-2-i)!(\nu-2+n-j)!}{(\nu-1-i-j)!i! j!} b^i a^j \Biggr] \notag\\
&= 
\sum_{j=0}^{\nu-2} (\nu-1-j)\frac{(\nu-1)!(\nu-2+n-j)!}{(\nu-1-j)! j!}
a^j \notag\\ 
&\quad + \sum_{i=1}^{\nu-2} \sum_{j=0}^{\nu-1-i}(\nu-1-j)  
\frac{(\nu-1-i)!(\nu-2+n-j)!}{(\nu-1-i-j)!i! j!} b^i a^j\,.
\label{eq:T2}
\end{align}
In the second step, we have shifted $i\to i-1$ in the first sum.  In
the third step, we have extracted the $i=0$ term of the second sum and
merged the remaining sums.  In the fourth step, we have shifted $j\to
j-1$ in the first sum in square brackets.  In the fifth step, we have
extracted the $j=\nu-1-i$ term of the first sum and the $j=0$ term of
the second sum in square brackets and merged the remaining sums.  In
the last step, we have written $1/(\nu-j-2)!$ as
$(\nu-1-j)/(\nu-1-j)!$ in the first sum.  Also, we have observed that
$ij+(\nu-1)(\nu-1-i-j)=(\nu-1-i)(\nu-1-j)$ and merged the three terms
in square brackets into a single sum.  

The result of eq.~\eqref{eq:T2} suggests to merge the two sums since
the first sum is just the $i=0$ term in the second sum, up to a missing $i=0$, $j=\nu-1$ term.  However, 
such a merge needs to be done with care as the extra term 
is formally undefined for $n=0$.  
We therefore first split $\nu-1-j$ as
$(\nu-1+n-j)-n$ in both sums and substitute $T_{\nu,n}$ in
eq.~\eqref{Sigma}.  We switch the order of the summations and first
perform the sum over $n$.  Considering only the terms that depend on
$n$, the sum to be evaluated is
\begin{align}
  &\sum_{n=0}^k (n+1)_{k-n}\bigl[(\nu-1+n-j)!-n(\nu-2+n-j)!\bigr]\notag\\
  &\qquad
  =\sum_{n=0}^k(n+1)_{k-n}(\nu-1+n-j)!-\sum_{n=1}^k(n)_{k-n+1}(\nu-2+n-j)!
  \notag\\
  &\qquad
  =\Biggl[\sum_{n=0}^k-\sum_{n=0}^{k-1}\Biggr](n+1)_{k-n}(\nu-1+n-j)!
  =(\nu-1+k-j)!\,,
\end{align}
where in the second step we have used $n(n+1)_{k-n}=(n)_{k-n+1}$ and
observed that the $n=0$ term does not contribute to the second sum,
and in the third step we have shifted $n\to n+1$ in the second sum so
that only the $n=k$ term remains.  Using this result we obtain
\begin{align}
  \Sigma_{\nu,k}&=
  \sum_{j=0}^{\nu-2} \frac{(\nu-1)!(\nu-1+k-j)!}{(\nu-1-j)! j!} a^j 
  + \sum_{i=1}^{\nu-2} \sum_{j=0}^{\nu-1-i}
  \frac{(\nu-1-i)!(\nu-1+k-j)!}{(\nu-1-i-j)!i! j!} b^i a^j\notag\\
  &=\sum_{i=0}^{\nu-2} \sum_{j=0}^{\nu-1-i}
  \frac{(\nu-1-i)!(\nu-1+k-j)!}{(\nu-1-i-j)!i! j!} b^i a^j
  -k!a^{\nu-1}\notag\\
  &=  \sum_{i=0}^{\nu-1} \sum_{j=0}^{\nu-1-i}
  \frac{(\nu-1-i)!(\nu-1+k-j)!}{(\nu-1-i-j)!i! j!} b^i a^j 
  - k! a^{\nu-1} 
  - \frac{(\nu-1+k)!}{(\nu-1)!} b^{\nu-1}  \,,
\end{align}
where in the second step we have merged the sums and corrected for the
$i=0$, $j=\nu-1$ term, and in the last step we have extended the sum
over $i$ up to $\nu-1$ and corrected for the $i=\nu-1$, $j=0$ term.
Looking back at eq.~\eqref{Snuk} and using
$(\nu-1+k)!/(\nu-1)!=(\nu)_k$ this yields
\begin{align}
  \Sigma_{\nu,k} &= S_{\nu,k} - k! a^{\nu-1} - (\nu)_k b^{\nu-1} \,,
\end{align}
which completes the proof.

\section{Error estimation for the random matrix simulations}
\label{App:err}

When using reweighting the ensemble average \eqref{fgw} can be written as $\bar z=\bar x/\bar y$, and the error on $\bar z$ is given by the usual error propagation formula
\begin{align}
\sigma_{\bar z} =  |\bar z|\sqrt{\frac{\sigma_{\bar x}^2}{{\bar x}^2} + \frac{\sigma_{\bar y}^2}{{\bar y}^2} - 2 \rho_{xy}\frac{\sigma_{\bar x}}{\bar x}\frac{\sigma_{\bar y}}{\bar y}}\,,
\label{stderr}
\end{align}
where $\sigma_{\bar x} = \sigma_x/\sqrt{N}$ and $\sigma_{\bar y}=\sigma_y/\sqrt{N}$ are the standard errors of $\bar x$ and $\bar y$, $\sigma_x$ and $\sigma_y$ are the square roots of the sample variances of $x$ and $y$, $\rho_{xy}$ is the correlation coefficient of $x$ and $y$, and $N$ is the sample size.

In the quenched simulation all configurations are independent and the standard errors are computed using the total sample size $N$. For the phase-quenched and sign-quenched
simulations the ensembles are generated using a Metropolis algorithm,
and the autocorrelations in the Markov chain have to be taken into
account by modifying the standard errors to
\begin{align}
\sigma_{\bar x} = \sqrt{\frac{2 \tau_{\text{int},x}}{N}}\,\sigma_x\,, \qquad
\sigma_{\bar y} = \sqrt{\frac{2 \tau_{\text{int},y}}{N}}\,\sigma_y\,,
\end{align}
where $\tau_{\text{int},x}$ and $\tau_{\text{int},y}$ are the
integrated autocorrelation times for $x$ and $y$ \cite{Madras:1988ei}.
In eq.~\eqref{stderr} the correlation coefficient $\rho_{xy}$ is
computed over the complete sample, and it is assumed that $\rho_{xy}$
is not affected by different autocorrelation times for $x$ and $y$.

\sloppypar
\bibliography{biblio} 
\bibliographystyle{jbJHEP}

\end{document}